\documentclass[journal]{IEEEtran}
\usepackage{cite}

\ifCLASSINFOpdf
  \usepackage[pdftex]{graphicx}
\else
\fi
\usepackage[cmex10]{amsmath}
\usepackage{algorithm}
\usepackage{algorithmic}

\makeatletter
\newcommand\fs@norules{\def\@fs@cfont{\bfseries}\let\@fs@capt\floatc@ruled
  \def\@fs@pre{}%
  \def\@fs@post{}%
  \def\@fs@mid{\kern3pt}%
  \let\@fs@iftopcapt\iftrue}
\makeatother
\floatstyle{norules}
\restylefloat{algorithm}

\usepackage{array}
\usepackage{url}

\usepackage{amssymb}
\usepackage{multirow}
\usepackage{arydshln}

\usepackage{bm}
\newcommand{\mbf}{\mathbf}
\def\app#1#2{%
  \mathrel{%
    \setbox0=\hbox{$#1\sim$}%
    \setbox2=\hbox{%
      \rlap{\hbox{$#1\propto$}}%
      \lower1.1\ht0\box0%
    }%
    \raise0.25\ht2\box2%
  }%
}

\newcommand{\Lf}{\mbf{L}_{off}}
\newcommand{\A}{\bm\alpha}
\newcommand{\Ta}{\bm\tau_a}

\newcommand{\Bp}{\mbf{B}_p}
\newcommand{\Bg}{\mbf{B}_g}
\newcommand{\Eg}{\bm\epsilon_g}
\newcommand{\Rg}{\bm\rho_g}
\newcommand{\Ld}{\mbf{L}_D}
\newcommand{\Lu}{\mbf{L}_U}
\newcommand{\N}{\mathcal{N}}
\newcommand{\R}{\mathcal{R}}

\begin{document}
%
\title{Improved Background Estimation for Gas Plume Identification in Hyperspectral Images}
%
%
%

\author{Scout~Jarman,~\IEEEmembership{Member,~IEEE,}
        Zigfried~Hampel-Arias,~\IEEEmembership{Member,~IEEE,}
        Adra~Carr,~\IEEEmembership{Member,~IEEE,}
        and~Kevin~R.~Moon,~\IEEEmembership{Member,~IEEE}
\thanks{Manuscript recieved DD Month YYYY, revised DD Month YYYY; accepted DD Month YYYY. Date of publication DD Month YYYY; date of current version DD Month YYYY. (\textit{Corresponding author: Scout Jarman)}}%
\thanks{Scout Jarman and Kevin R. Moon are with the Deptartment of Mathematics and Statistics at Utah State University, Logan, UT 84322 USA (email: scout.jarman@usu.edu; kevin.moon@usu.edu).}
\thanks{Zigfried Hampel-Arias and Adra Carr are with the Intelligence and Space Research division at Los Alamos National Laboratory, Los Alamos, NM 87545 USA (email: zhampel@lanl.gov; acarr@lanl.gov).}}

%
%

\markboth{IEEE transactions on geoscience and remote sensing, VOL. XX, 202X}%
{Jarman \MakeLowercase{\textit{et al.}}: Improved Background Estimation for Gas Plume Identification in Hyperspectral Images}
%



\maketitle

\begin{abstract}
Longwave infrared (LWIR) hyperspectral imaging can be used for many tasks in remote sensing, including detecting and identifying effluent gases by LWIR sensors on airborne platforms.
Once a potential plume has been detected, it needs to be identified to determine exactly what gas or gases are present in the plume.
During identification, the background underneath the plume needs to be estimated and removed to reveal the spectral characteristics of the gas of interest.
Current standard practice is to use ``global" background estimation, where the average of all non-plume pixels is used to estimate the background for each pixel in the plume.
However, if this global background estimate does not model the true background under the plume well, then the resulting signal can be difficult to identify correctly.
The importance of proper background estimation increases when dealing with weak signals, large libraries of gases of interest, and with uncommon or heterogeneous backgrounds.
In this paper, we propose two methods of background estimation, in addition to three existing methods, and compare each against global background estimation to determine which perform best at estimating the true background radiance under a plume, and for increasing identification confidence using a neural network classification model.
We compare the different methods using 640 simulated plumes.
We find that PCA is best at estimating the true background under a plume, with a median of 18,000 times less MSE compared to global background estimation.
Our proposed K-Nearest Segments algorithm improves median neural network identification confidence by 53.2\%.
\end{abstract}


%
\IEEEpeerreviewmaketitle

\section{Introduction}

\begin{figure*}
    \centering
    \includegraphics[width=\linewidth]{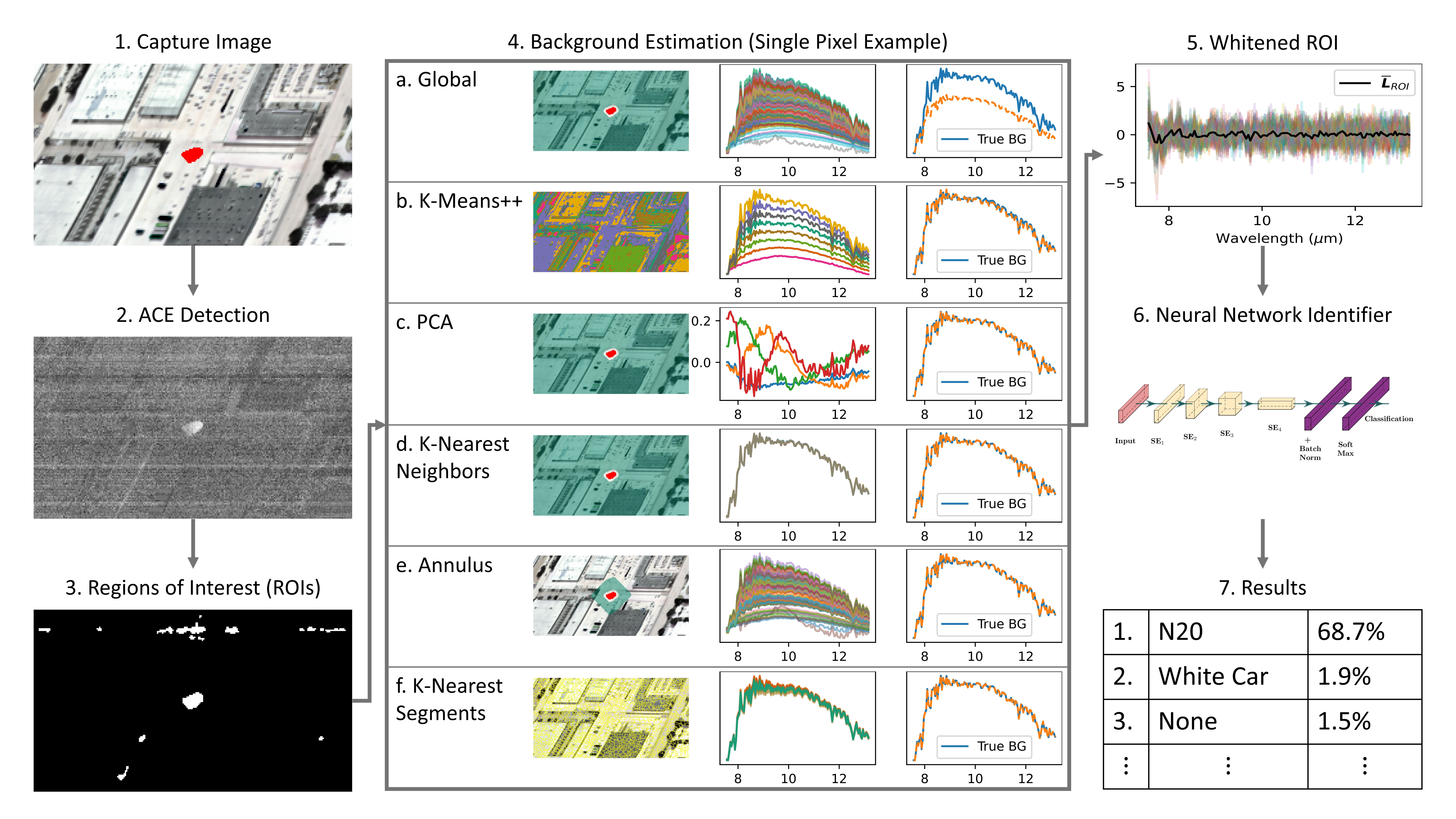}
    \caption{Example of gas plume identification process: 1. False color LWIR image from the Los Angeles Basin with a simulated N$_2$O plume (shown in red). 2. ACE detection map for a reference N$_2$O signal at 280K and a globally estimated background. 3. Automatically created regions of interest (shown in white). 4. The various background estimation strategies once a ROI has been created, where each method is applied to a single example pixel in the ROI. Green shading indicates which pixels are being used in the estimation process with the exception of K-means++, which shows the cluster assignments, and K-nearest segments which shows the image segmentation boundaries. 5. Individually whitened ROI pixels, and the final averaged whitened spectral signature (black). Example was whitened using the K-nearest segments background estimation with 32 minimum-pixels, complete linkage, and background target separation. 6. A basic overview of the neural network identifier architecture. 7. Neural network prediction confidences. For reference, the top three predictions using global background estimation are None (4.4\%), Calcium Sulfate (4.1\%), and N$_2$O (3.5\%).}
    \label{fig:Overview}
\end{figure*}

\IEEEPARstart{H}{yperspectral} imaging (HSI) is increasingly useful and versatile, with applications ranging from environmental monitoring and medical diagnosis to national security \cite{Stuart2019HyperspectralII, Pallua2021NewPO, Yuen2010AnIT, Khan2018ModernTI}.
HSI captured in the long-wave infrared (LWIR) electromagnetic spectrum is particularly useful in gas plume analysis due to many gases having unique ``finger prints'' in LWIR wavelengths \cite{Hulley2016HighSR, Burr2006OverviewOP}.
Gas plume analysis can be broken into three main tasks: detection, identification, and quantification (the first two steps are shown in Figure \ref{fig:Overview}).

Detection is the process of determining which pixels in an image contain a gas of interest, typically with the use of a matched filter, adaptive coherence estimator, or similar detectors \cite{Manolakis2014DetectionAI, Manolakis2002DetectionAF}.
Once a potential gas plume, or region of interest (ROI), has been created from the detection stage, a separate model is used to identify the gas or gases in the plume.
This identification step helps to reduce false positives from the detection step, and to distinguish between similar spectral signatures \cite{Truslow2016PerformanceMF}.
Popular identification models include Bayesian model averaging, step-wise linear regression, and more recently neural network based models \cite{Burr2002ChemicalIU, Hoeting1999BayesianMA, Pogorzala2005GasPS, Li2019DeepLF, Klein2023HyperspectralTI}.
Lastly, once the plume has been identified, the quantity and temperature of the gas plume can be estimated \cite{Idoughi2016BackgroundRE, Hayden1997RemoteTG}.
This analysis workflow has allowed gas plume analysis to be applied to application areas such as industrial and environmental monitoring and disaster response \cite{Manolakis2019LongwaveIHPPC, Manolakis2014LongWaveIH, Buckland2017TrackingAQ}.

The development of detection, identification, and quantification algorithms depends on understanding the radiative transfer equation for gas plumes.
More detail about the radiative transfer equation will be given in Section \ref{sec:Background}.
For simplicity, it is common to use the additive model (or signal plus noise model) instead of the full radiance equation.
The additive model states that for a gas plume with one gas, the observed radiance for any pixel is modeled as
\begin{equation}
    \label{eq:additive model}
    \mbf L = \Lf + \psi\bm s,
\end{equation}
where $\Lf$ is the ``off plume" or background radiance, $\psi$ is a signal strength scalar, where $\psi=0$ indicates a pixel with no plume present, and $\bm s$ is the shape of the gas or target, more specifically the gas's absorption coefficient.
To isolate the shape of the gas $\bm s$ scaled by $\psi$, we must estimate and subtract the background $\Lf$.

In LWIR HSI, several factors contribute to variation of the background signals, including different background surface materials, temperatures, sensor noise, and natural variability of spectral signatures.
As a result, it is common to assume that $\Lf$ follows a multivariate normal distribution with mean $\bm\mu_g$ and covariance $\Sigma_g$.
This assumption is the foundation of many algorithms including the matched filter, adaptive coherence estimator, step-wise linear regression, and Bayesian model averaging \cite{Manolakis2014DetectionAI}.
The assumption also allows for spectral ``whitening" which suppresses background effects and highlights important spectral features from the gas.
It is described as
\begin{equation}
\label{eq:white}
    \widetilde{\mbf L} = \Sigma_g^{-1/2}(\mbf L - \bm\mu_g).
\end{equation}
Since $\bm\mu_g$ and $\Sigma_g$ are often estimated using all pixels in a scene, this approach is referred to as a ``global" algorithm.

Though the normality assumption is useful to develop detection and identification algorithms, real images typically do not satisfy the assumption \cite{Burr2006OverviewOP}.
When these assumptions are not satisfied, detection algorithm performance suffers \cite{Nasrabadi2014HyperspectralTD}, prompting research in alternative background modeling strategies for detection \cite{Matteoli2014AnOO}.
However, background modeling for identification purposes has been less studied.

This paper focuses on improving neural network identification performance of whitened gas signatures.
Since the whitened signatures are standardized before being passed into the neural network, accurate background estimation plays a critical role in gas identification.
We consider three non-spatial methods for estimating background radiance, namely principal components analysis, K-means clustering, and K-nearest neighbors, along with two spatial methods, namely the annulus method and K-nearest segments (see Figure \ref{fig:Overview} for analysis pipeline and methods).
These methods are evaluated by comparing background radiance estimation performance and gas identification performance, benchmarked against standard global background estimation methods.

This paper has three main contributions to the field of background estimation.
First, we have adapted and directly compared existing background estimation techniques for gas plume identification.
Second, we proposed two new methods for background estimation, namely K-nearest neighbors and K-nearest segments.
Third, we simulated 640 plumes allowing us to analyze background estimation performance across different signal strengths and over many different backgrounds.

We structure our results as follows:
Section \ref{sec:Background} provides additional detail on the radiative transfer equation, highlighting the importance and difficulty of background estimation.
Section \ref{sec:Methods} demonstrates how existing methods can be extended to identification, along with our proposed methods.
Additionally, a description of the neural network identifier and our plume simulation algorithm is given.
Section \ref{sec:Results} compares the performance of each background estimation method when applied to both background radiance estimation and identification confidence.
We also include an analysis of hyperparameter tuning and sensitivity for each method.
Lastly, Section \ref{sec:Conclusions} summarizes conclusions and future works.

\section{Background}
\label{sec:Background}

The development of many detection and identification algorithms begins with the radiative transfer equation.
This equation describes what a sensor ``sees" and specifically how the presence or absence of a gas plume influences the observed radiance.
Understanding the equation will demonstrate the importance and difficulty in estimating the background radiance, and motivate the development of background estimation strategies.

A full description of the radiance transfer equation is found in \cite{Manolakis2019LongwaveIHPPC}.
It is described by the background, or ``off plume" radiance, and the effects from the gas.
These terms are described as:
\begin{equation}
    \Lf=\Lu + \Ta(\Eg\Bg + \Rg\Ld),
\end{equation}
\begin{equation}
\label{eq:Lon}
    \mbf L = \Lf + n_c\A\Ta(\Bp - \Eg\Bg - \Rg\Ld),
\end{equation}
where $\Lu, \Ld$, and $\Ta$ are the atmospheric upwelling, downwelling, and transmission respectively.
$\Eg, \Rg$, and $\Bg$ are the background surface emissivity, reflectance, and blackbody radiance at temperature $T_g$.
When a plume is present, $n_c, \A$, and $\Bp$ are the gas plume's concentration pathlength, absorption coefficient, and blackbody radiance at temperature $T_p$.
When a plume is not present $n_c=0$ and $\mbf L=\Lf$.
Each term except for $n_c$ is a vector where each element corresponds to an observed wavelength $\lambda$, and element wise multiplication is used.

Equation \ref{eq:Lon} demonstrates the importance of background estimation.
The observed signal $\mbf L$ is primarily composed of the background $\Lf$, and the slight differences caused by the interaction of the gas with the background radiance passing through it.
Therefore, if it is possible to estimate the background terms, the effects the gas has on the observed signal can be isolated.
Once the gas's effect is estimated, detection and identification of the gas can proceed by comparing the estimated signal to a library of known gas signals.

Equation \ref{eq:Lon} is also the foundation of the simplified additive model.
In the additive model from Equation \ref{eq:additive model}, the $\Lf$ term remains the same, while the target $\bm s$ represents $\A\Ta$, and the signal strength $\psi$ represents $n_c(\Bp - \Eg\Bg - \Rg\Ld)$.
The additive model is an approximation since $\psi$ should be a vector, and there are background effects, namely $\Eg, \Bg$, and $\Rg$, that are present in both the $\Lf$ and $\psi$ terms.
However, it is a useful model to develop detection and identification algorithms.

\subsection{Detection and Identification}

The purpose of the detection stage is to determine if a known signal of interest is present in the image.
Though there exist many detection algorithms, the adaptive coherence estimator (ACE) is primarily discussed here (for a more in-depth description of ACE and other detection algorithms, see \cite{Manolakis2014LongWaveIH} and \cite{Manolakis2009IsTA}).
Using Equations \ref{eq:additive model} and \ref{eq:white}, $\widetilde{\bm s} = \Sigma_g^{-1/2}\bm s$ for a signal of interest $\bm s$, and assuming that $\Lf\sim N(\bm\mu_g, \Sigma_g)$, the ACE detection score is defined as:
\begin{equation}
    ACE(\mbf L, \bm s) = \frac{(\widetilde{\bm s}^T\widetilde{\mbf L})^2}{(\widetilde{\bm s}^T\widetilde{\bm s})(\widetilde{\mbf L}^T\widetilde{\mbf L})}.
\end{equation}
Thresholding and region merging is used to automatically create ROIs.

ACE has a constant false alarm rate based on the chosen threshold.
The purpose of identification algorithms, such as Bayesian model averaging and step-wise least squares, is to reduce the false alarm rate and determine whether a mixture of gases is present.
This work focuses on the single gas identification case, specifically using a neural network.

The neural network is trained as a classifier to take an input whitened signature and to determine its identity from a provided spectral library.
Since a ROI contains many pixels, the average whitened pixel, or superpixel, is found as
\begin{equation}
    \overline{\mbf L}_{ROI} = \frac{1}{n}\sum_{i=1}^n \widetilde{\mbf L}_{i}.
\end{equation}
This is then standardized before being passed as input to the neural network identifier (NNI).
The NNI is trained on millions of generated examples of known gas signatures $\widetilde{\bm s}$ and provides a determination of which $\widetilde{\bm s}$ a given $\overline{\mbf L}_{ROI}$ most resembles.
More details on the NNI are in Section \ref{sec:NNI}

The whitening transformation is based in statistics, namely ``spherizing" the general multivariate Gaussian distribution to be a standard normal multivariate Gaussian distribution, as seen in Equation \ref{eq:white}, where $\bm \mu_g$ is subtracted from $\mbf L$.
However, one benefit of using a NNI is that it does not enforce or require any strict distributional assumptions on the signal it is trying to identify.
This is of particular benefit given that it has been shown that HSI do not satisfy the normality assumption \cite{Nasrabadi2014HyperspectralTD, Matteoli2014AnOO, Burr2006OverviewOP, Carlotto2005ACA, Marden2004UsingEC, Matteoli2013ModelsAM}.
We therefore investigate the whitening transform from its relation to Equation \ref{eq:Lon}, instead of its statistical construction.

Based on Equations \ref{eq:additive model} and \ref{eq:white}, we try to isolate the gas effect $\bm s$ by removing the effect of the background $\Lf$.
Since the global approach assumes that $\Lf$ is normally distributed, it subtracts the maximum likelihood estimate of $\Lf$, which is $\bm \mu_g$.
Though $\bm\mu_g$ is in theory a good estimate of $\Lf$ overall, it is not likely to be a good estimate of any specific $\Lf$ in a ROI, not least because the assumption is not satisfied.
Having a poor estimate of $\Lf$ will lead to an $\overline{\mbf L}_{ROI}$ that may not match well the true $\widetilde{\bm s}$, leading the NNI to have less confidence in its prediction of what gas is present, or misidentify the gas entirely.
This problem is further exacerbated when considering complex scenes, like urban environments, where plumes are likely to pass over multiple unique backgrounds \cite{Jarman2023EnsembleSF}.

We propose to investigate better ways to estimate $\Lf$ that do not depend on the normality assumption, and that use additional information available in an image.
For the purpose of isolating the effect that $\Lf$ estimates have on identification performance, we still use the globally estimated $\Sigma_g$ in the whitening process.
Further research on alternative whitening approaches should include a study of different estimates of $\Sigma_g$.

\section{Methods}
\label{sec:Methods}

There has been much research in the field of background modeling, particularly for detection \cite{Matteoli2014AnOO}.
However, there is a notable distinction between detection and identification.
Detection is applied to each pixel independently, while identification takes the aggregate of all individually whitened pixels in a given ROI.
This means there is a spatial component with identification since the ROI is a collection of nearby, possibly heterogeneous, pixels.
Consequentially, many background modeling methods previously applied to detection will need to be modified for the purpose of identification.

We tested five methods, and compared each against global background estimation.
There are two main categories of background estimation methods, spatial and non-spatial.
Spatial methods use the spatial structure of an image, while non-spatial methods treat each pixel independent of location.
Two of the five tested methods are new, and, to the best of our knowledge, have not been applied to gas analysis in LWIR HSI.
The remaining three methods have been used in gas plume detection and quantification, but not in identification.

The tested methods are K-means++ clustering (KMeans), principal components analysis (PCA), the annulus method (Annulus), and our new proposed methods of K-nearest neighbors (KNN) and K-nearest segments (KNS).
Figure \ref{fig:Overview} gives a brief visual description of how each method works.

For each method, the ROI is morphologically dilated four times in succession to create a guardrail of pixels not to be included in the background estimation.
This is done so that weak gas signals on the edge of a plume that fail to reach the detection threshold, do not contaminate the background estimates.
We will refer to the set of ROI pixels as $\R$, and the set of non-ROI pixels, excluding pixels in the guard rail, as $\N$.

Global background estimation does not have any hyperparameters, which makes it an attractive option.
Every other method that we consider has one hyperparameter to tune except for KNS having three.
The choice of hyperparameter, often simply $k$ for the number of clusters or number of pixels to use, will alter the performance of each background estimate.
An exploration of the optimal hyperparameters and hyperparameter sensitivity is discussed in Section \ref{sec:hyperparameter}.

\subsection{Global}

Global background estimation is the baseline method against which each of the following methods is compared.
The Global estimate is found as the average of the pixels in $\N$.
This single background estimate is calculated and used for each pixel in $\R$.

The primary benefit of using the Global estimate is that it is fast and easy to compute and has been shown to be successful in some circumstances even though the normality assumption does not hold \cite{Manolakis2009IsTA}.
However, for plumes that traverse rare backgrounds, or when the normality assumption is particularly mismatched, the Global estimate may not estimate $\Lf$ well.
Furthermore, a plume that traverses multiple backgrounds may produce a noisy whitened signal when using the Global estimate since the same background is subtracted from each pixel in $\R$.

\subsection{K-Means++}

K-means clustering is a popular clustering algorithm \cite{Hartigan1979AKC}.
We specifically use the K-means++ algorithm, which has improved seeding to speed up convergence, though we will simply refer to it as KMeans \cite{Arthur2007kmeansTA}.
KMeans, and related clustering algorithms like Gaussian mixture models, have been used to improve detection performance \cite{Matteoli2014AnOO, Funk2001ClusteringTI}.
It has also been used to model background radiance for the purpose of quantification \cite{Idoughi2016BackgroundRE}.

In our implementation, the $k$ clusters are fit on $\N$.
For each pixel in $\R$, the Euclidean distance between each of the $k$ cluster centers is calculated.
The background estimate for the pixel is the cluster center corresponding to the smallest Euclidean distance.

In contrast to the Global estimate, this method finds a set of $k$ different backgrounds.
This is particularly helpful when a plume traverses many backgrounds, as each one should be taken into account.
KMeans may be less beneficial for the case of rare backgrounds, as rare backgrounds may be included in a larger cluster, which the choice of $k$ will impact.
Furthermore, as seen in Figure \ref{fig:Overview}, KMeans tends to create cluster centers according to apparent surface temperature, and not emissivity characteristics.
If a particular background has a unique spectral shape, the cluster center may not characterize this well.

\subsection{Principal Components Analysis}

PCA is a popular dimensionality reduction method that works by finding a set of linearly independent vectors aligned with the direction of the most variance in the data\cite{Jolliffe2002PrincipalCA}.
These principal components are sorted from most variability explained to least.
In gas plume analysis, PCA is often used to define a subspace that describes the background materials.
This approach has been used to improve both detection and quantification of gases \cite{Manolakis2001HyperspectralST, Hayden1996DeterminationOT, Hayden1997RemoteTG}.

The principal components are found using $\N$.
Then, each pixel in $\R$ is projected onto the subspace spanned by the principal components.
Since PCA is fit on $\N$, the projection should in theory be a sufficiently faithful reconstruction of $\Lf$ \cite{Hayden1997RemoteTG}.
The projection is used as the estimate of $\Lf$ on a per pixel basis.
This method is again fast and simple, though the user must choose how many principal components to use.

It should be noted that it is best practice to ignore the channels where a gas of interest is most prominent \cite{Hayden1997RemoteTG}.
However, we have the more difficult challenge where we assume no \textit{a priori} information about the gas in the ROI, and therefore cannot ignore any channels.
In practice, if a ROI has detections for several different materials, a different PCA background could be estimated for each detected material.

\subsection{K-Nearest Neighbors}

KNN is a classic machine learning algorithm that can be used for both classification and regression \cite{Bao2016IntroductionTM}.
KNN has been used in other fields of hyperspectral analysis such as image classification \cite{Ma2010LocalML, Huang2015SpectralSpatialHI, Cui2015ClassDependentSR, Guo2017KNearestNC}.
It has also been used for automatic bandwidth selection for kernel density estimation for background modeling, but, to the best of our knowledge, it has not been used solely for background estimation \cite{Matteoli2014BackgroundDN, Matteoli2014ALA}.

For background estimation, $\N$ is used as the training dataset.
For each pixel in $\R$, the background is estimated as the average of the $k$ pixels in $\N$ that have the smallest Euclidean distance.
It should be noted that ``nearest" is defined in terms of spectral distance and not spatial distance.

KNN can theoretically be slow to calculate depending on the sizes of $\N$ and $\R$, however with modern computation power, particularly with GPU acceleration, it is not a practical limiting factor.
Similar to KMeans, KNN helps with plumes that traverse multiple backgrounds, since each pixel has its own background estimate.
It is also able to handle the case of rare backgrounds, as long as an appropriate $k$ is chosen, and the rare background is present outside of the guard rail of pixels.

KNN suffers from a similar problem as PCA, in that the Euclidean distance will increase by using wavelengths in which the gas signal is most prominent.
This problem is exacerbated for strong gases that have larger thermal contrasts or pathlength concentrations.
However, a plume strong enough to significantly disrupt Euclidean distance is likely strong enough to be easily detected and identified using a Global background.
In our study, we focus on weak plumes that do not significantly increase Euclidean distance.

In practice, if the number of gases of interest is small, then multiple backgrounds could be calculated by ignoring each gas's prominent wavelengths.
Additionally, other ROI's suspected to contain the same gas found in $\R$ should not be included in $\N$.

\subsection{Annulus Method}

Annulus is the first background estimation algorithm that includes spatial information to estimate the background and is therefore considered a local background estimation technique.
This approach has been used to make improved local detectors \cite{Acito2005AdaptiveDA, Matteoli2010ATO, Nasrabadi2014HyperspectralTD}.

For detection, an annulus is grown around a single pixel.
We extend this to a gas plume by successively dilating the ROI guardrail $k$ times, and removing the guardrail and ROI pixels from the calculations.
The background is then estimated as the average of all pixels in the annulus around the ROI.
Similar to the Global estimate, the annulus produces a single background estimate used for each pixel in $\R$.

In a small enough area, neighboring pixels should be spectrally similar to each other.
This can be exploited for background estimation since the clean pixels around a ROI should be of a similar background to the contaminated pixels in $\R$. 
However, if a plume is traversing many background materials, or the annulus is too large, the average may not be a good estimate of any of the non-dominant backgrounds. 
These shortcomings are the result of the annulus using only spatial information and being ignorant of spectral differences.

\subsection{K-Nearest Segments}

Each of the previous methods offer benefits over Global estimation, but also have shortcomings.
KNN is able to handle multiple and rare backgrounds in a plume, but suffers from plume contamination.
Annulus uses the spatial information from the image, but is unaware of the fact that backgrounds will also vary in space.
In \cite{Jarman2023EnsembleSF}, a ``fuzzy annulus" was created by using an ensemble of watershed segmentations, however the method was complex, computationally expensive, and hard to tune for many different plume morphologies.
We propose the KNS method that aims to exploit the benefits of each of the previous methods, while reducing the downsides of each, and is easier to tune for many different plume morphologies.

It begins by extracting more spatial information from the hyperspectral image using image segmentation.
Image segmentation is used to break the image into many different background regions of homogeneous pixels \cite{Grewal2022HyperspectralIS}, each treated individually.
Though there exist many image segmentation algorithms, we use watershed segmentation due to it being unsupervised and adapted for use with HSI \cite{Tarabalka2010SegmentationAC}.
Further investigation of how other segmentation algorithms affect performance is needed.

Let $S_\R$ be a set containing all segments that contain ROI pixels, and $S_\N$ be a set containing all segments that contain non-ROI pixels.
There exist many possible ways to measure the similarity, or distance, between all pairs of segments from $S_\R$ to $S_\N$.
To use all the information from all the pixels in a segment, we use linkage functions from hierarchical clustering to measure the similarity between pairs of segments \cite{Ding2002ClusterMA}.
We consider single, complete, and average linkage functions, defined respectively as:
\begin{equation}
    D_{s}(A, B) = \min_{a\in A, b\in B}d(a,b)
\end{equation}
\begin{equation}
    D_{c}(A, B) = \max_{a\in A, b\in B}d(a,b)
\end{equation}
\begin{equation}
    D_{a}(A, B) = \frac{1}{|A|\cdot|B|}\sum_{a\in A}\sum_{b\in B}d(a,b),
\end{equation}
where $A$ is a segment from $S_\R$, $B$ is a segment from $S_\N$, and $d(a,b)$ is the Euclidean distance between two spectral signatures, though other distance functions may be beneficial \cite{Deborah2015ACE, Jarman2024LocalBE}.
The single linkage $D_s$ is the shortest distance between all pairs of points in two segments.
Complete linkage $D_c$ is similarly the largest distance, and average linkage $D_a$ is the average of all pairs of distances.
The choice of linkage function is one of the three hyperparameters for KNS.

We let the hyperparameter $k$ represent the minimum number of pixels to be used to estimate the background, instead of the actual number of neighboring segments.
This is done because segments can be of drastically different sizes, leading to different computation times for similar $k$ values.
Thus $k$ is the second hyperparameter for KNS.

Consider a specific segment $S_r\in S_\R$.
Create a collection of non-ROI pixels $S_n$ by ordering the segments from $S_\N$ according to their similarity, and combining the most similar segments until there are at least $k$ pixels ($|S_n|\geq k$).
Let $\mbf L_i$ represent ``clean" pixels from $S_n$, and $\overline{\mbf L}_i$ represent ``contaminated" pixels from $S_r$.
A modified version of the additive model is used with $\overline{\mbf L}_i=\Lf +\bm\psi_i\mbf t$, where the signal strength is a vector and element-wise multiplication is used for $\bm\psi_i \mbf t$.
If we assume that $\bm\psi_i=\bm0$ for all pixels in $S_n$, we can model $\mbf L_i=\Lf$.
If we assume that all pixels in $S_r$ and $S_n$ have the same background $\Lf$, then one way to estimate the background is by solving
\begin{equation}
\label{eq:mse bg}
    \min_{\Lf} \frac{1}{|S_n|}\sum_{\mbf L_i\in S_n}||\mbf L_i - \Lf||^2.
\end{equation}
This is equivalent to finding the background $\Lf$ that minimizes the mean squared error (MSE) between the non-ROI pixels and the estimated background.
The solution is the average of the non-ROI pixels in $S_n$.

However, $\overline{\mbf L}_i$ is also modeled with $\Lf$.
The information contained in the contaminated pixels can be included into the background estimate by solving
\begin{equation}
\label{eq:bts}
    \begin{tabular}{rl}
        $\displaystyle \min_{\Lf, \bm\psi_i, \mbf t}$ & $\displaystyle \frac{1}{|S_n|}\sum_{\mbf L_i\in S_n} ||\mbf L_i-\Lf||^2 +$\\\\
         & $\displaystyle \frac{1}{|S_r|}\sum_{\overline{\mbf L}_i\in S_r} ||\overline{\mbf L}_i - (\Lf + \bm\psi_i\mbf t)||^2$\\\\
         $s.t.$ & $\psi_{ij}\geq0, 0\leq t_j\leq1, \forall j,$
    \end{tabular}
\end{equation}
where a second MSE term is included between the contaminated pixels and the additive model.

We call this process background-target separation (BTS).
This can be solved numerically using the L-BFGS-B algorithm \cite{Byrd1995ALM}, instead of iteratively as proposed in \cite{Jarman2024LocalBE}.
The constraints are added to enforce the target shape staying in $\mbf t$, and the non-negative scale values staying in $\bm\psi_i$.
The direction of the constraint of $\bm\psi_i$ can be switched to be non-positive if the plume is suspected of being in absorption, or cooler than the background material.
The third hyperparameter is whether to estimate the background using BTS from Equation \ref{eq:bts}, or to not include the contaminated pixels by using Equation \ref{eq:mse bg}.

Compared to KNN, the use of a linkage function should reduce the impact that the gas has on skewing the distance calculations between pairs of pixels.
Compared to Annulus, KNS uses more spatial information as segmentation allows KNS to use locally homogeneous regions of pixels that are separated according to changes in background materials.
Unlike Annulus, which is restricted to pixels nearby the ROI, KNS is able to search the entire image for similar segments, allowing it to handle the case of rare backgrounds similar to KNN.

This method is more computationally expensive than the others because it requires image segmentation, pairwise distances of segments, and solving a potentially large optimization with BTS, but the use of multiprocessing and GPU acceleration make it tractable.

\subsection{Neural Network Identifier}
\label{sec:NNI}

Similar to the ACE detection algorithm, the standard identification algorithms are based on the normality assumption and the additive model.
On the other hand, neural network based models need not enforce specific distributional or linear assumptions.
This makes them well suited to the task of gas identification, particularly when considering a large library of reference materials \cite{Klein2022QuantifyingUI, Klein2023HyperspectralTI, HampelArias20242DSR}.

\begin{figure}
    \centering
    \includegraphics[width=\linewidth]{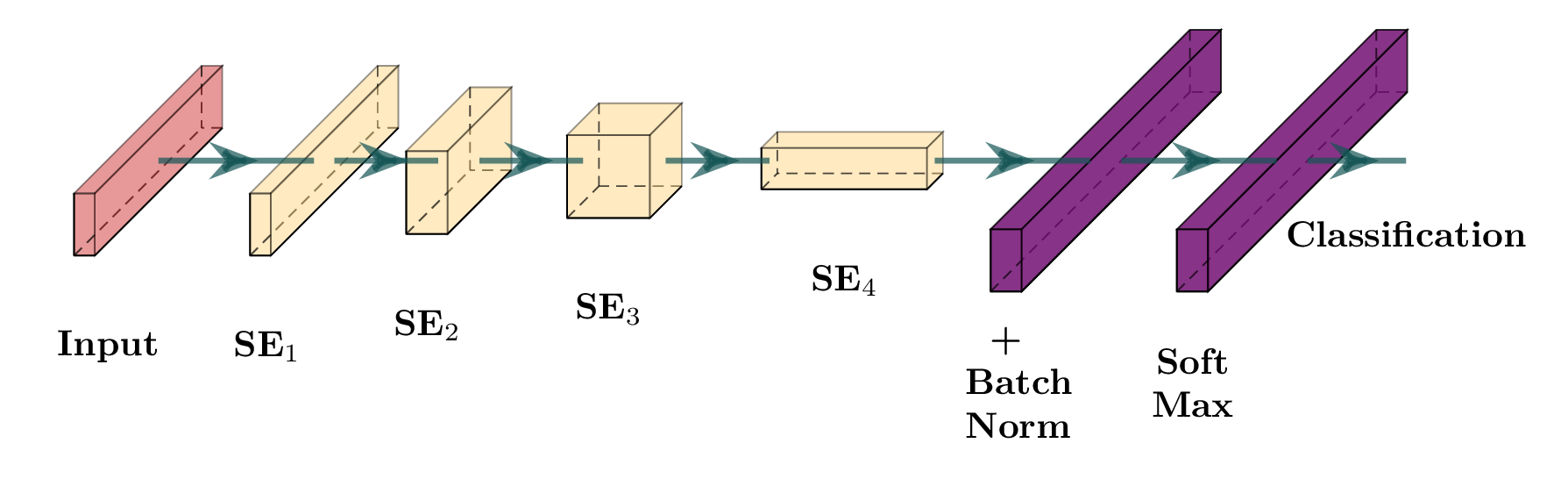}
    \caption{Diagram of the NNI architecture. The input is a 1-D vector, which is then passed through four squeeze-excite blocks, before being passed into a batch norm layer, and a final softmax layer for classification.}
    \label{fig:simplenet}
\end{figure}

The NNI used here is extended from \cite{Klein2022QuantifyingUI}.
The network is based on a relatively simple architecture that uses 1-D convolutional ``squeeze-excite" (SE) blocks to efficiently reduce data representation sizes, and create a rudimentary attention mechanism \cite{HasanPour2016LetsKI, Hu2017SqueezeandExcitationN}.
Figure \ref{fig:simplenet} gives a visualization of the architecture of our NNI.
The input is the 128 channel whitened signature, which is passed into four successive SE blocks with filter sizes of 512, 256, 128, and 64.
The final SE block gets flattened, passed into a batch norm layer, a 30\% dropout layer, then a final linear layer.

The final linear layer produces an output strength for each class, where a larger value indicates the model is more confident that the given class is correct.
Applying the softmax to the response strengths converts it into a pseudo probability, which is interpreted as identification confidence.
Typically, the material with the highest identification confidence is selected.

The size of the final linear layer depends on how many materials are of interest.
We use a library of reference signatures produced by Los Alamos National Laboratory, Pacific Northwest National Laboratory, and commercial sources.
Our final output layer contains approximately 10,500 materials, where only about 2,500 are gases.
Though this is a general model that aims to identify gases and solids, we are focused on how to improve this model's gas identification performance.

To train the model, example spectral signatures are generated that mimic what would be seen in the real world, using a modified form of Equation \ref{eq:Lon}:
\begin{equation}
    \begin{tabular}{rl}
        $ \mbf L_{train} =$ & $n_c\A\Ta(\Bp-\Eg\Bg-\Rg\Ld)$\\
         & $+(1-n_c\A)(\mbf L_{sp}-\Lf)$.
    \end{tabular}
\end{equation}
The second line adds noise to the training signal by simulating imperfect background subtraction.
For generating a sample from a given real world image, the terms $\Ta, \Ld, \Lu$ can be estimated. 
The terms $\Eg, \Rg, \Bg$ can be estimated from the global mean $\bm\mu_g$ using standard temperature-emissivity separation techniques (TES) \cite{Manolakis2019LongwaveIHPPC}, and the sample imperfect background $\mbf L_{sp}$ is computed as the average of a random rectangle of pixels.
Library reference signals for a gas at a certain temperature are used for $\A$ and $\Bp$.
A large or small $n_c$ can then be chosen to create a strong or weak gas signal, teaching the network to learn how to identify gases with varying levels of signal to noise.

This method of signature generation allows the creation of a dynamic training generator, as opposed to the traditional fixed, static datasets typically used to train neural networks.
For each epoch, 20 images are randomly selected from a pool of 4000.
Each cube will be used to generate 12,000 example signatures, for a total of 240,000 training samples per epoch.
The model is then trained for 100 epochs with a minibatch size of 64, using an Adam optimizer \cite{Kingma2014AdamAM}, and the binary cross entropy loss function.

\subsection{Gas Plume Simulation}

To evaluate and compare the performance of each method, plumes with a known background are needed.
Unfortunately such ground truth data is generally difficult to collect and unavailable.
We therefore simulate, and implant gas plumes into existing images, creating a large dataset of testing images.

Ten images from the Los Angeles Basin in California were used for simulation.
These images, and the preprocessing steps used, are similar to those found in \cite{Buckland2017TrackingAQ}.
These images have a pixel resolution of 128x2600, with 128 spectral channels ranging from 7.56 $-$ 13.16$\mu$m.
A Gaussian plume dispersion model, with an added random wind component, was used to generate the shape and concentration of the plume \cite{Holmes2006ARO}.

The time averaged plume concentration was scaled between zero and one such that one represents the maximum plume density and zero represents no plume.
The relative plume density is used to linearly model the plume's temperature and pathlength concentration $n_c$.
Each plume has a minimum temperature of 280K, that approached an ambient temperature of around 290$-$310K, meaning the plume signals were almost always in absorption.
TES and atmospheric estimation techniques were used to estimate non-gas related terms in Equation \ref{eq:Lon} on a per-pixel basis \cite{Manolakis2019LongwaveIHPPC}.

\begin{figure}
    \centering
    \includegraphics[width=\linewidth]{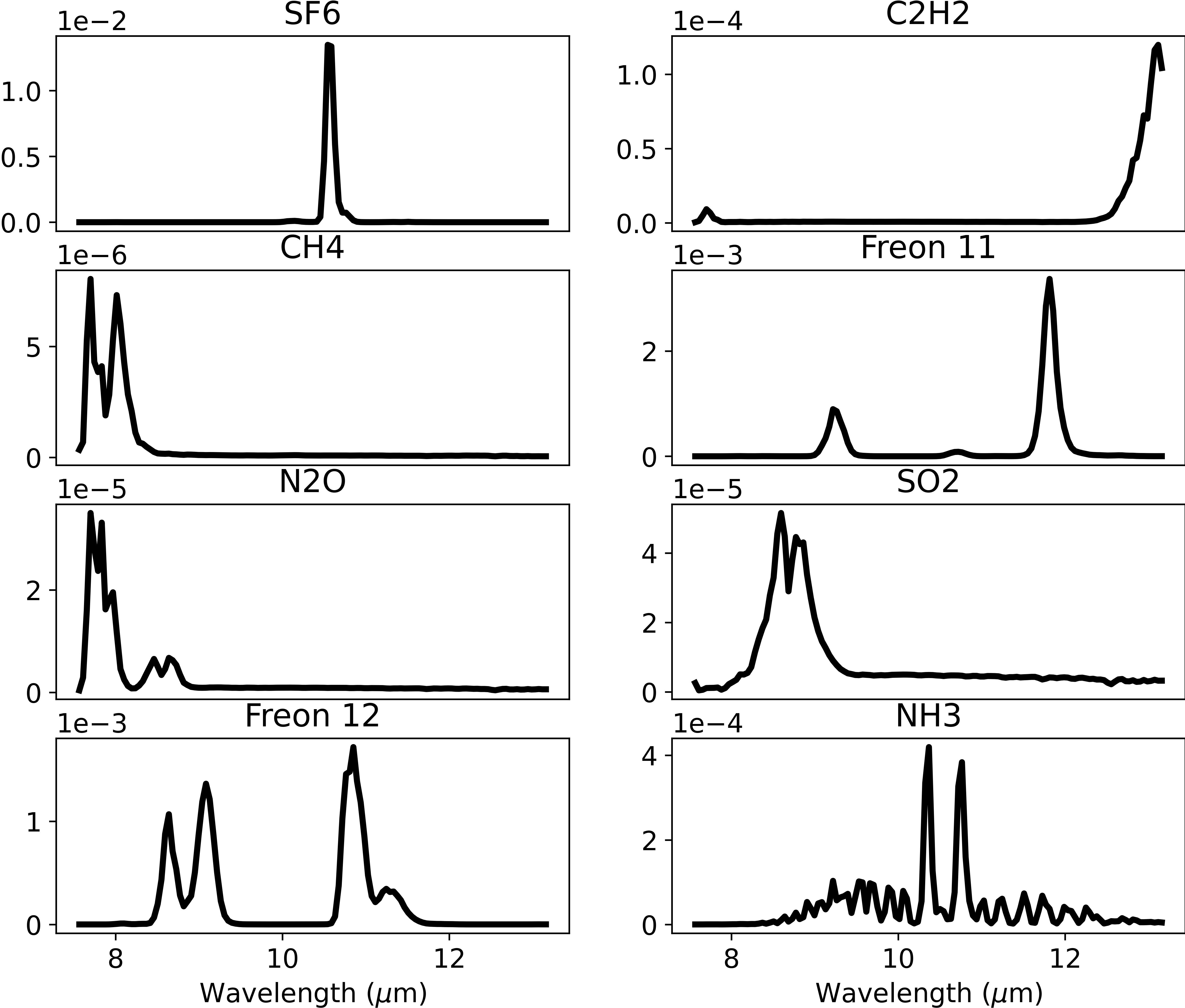}
    \caption{Library reference absorption signatures ($\A$) used to simulate gas plumes.}
    \label{fig:library sig}
\end{figure}

Figure \ref{fig:library sig} shows the reference library gas absorption signatures $\A$ used to simulate our plumes.
They are sorted with relatively narrow feature gases first, namely SF$_6$, C$_2$H$_2$, CH$_4$, Freon 11, and N$_2$O, followed by the wider feature gases SO$_2$, Freon 12, and NH$_3$.

As seen in Figure \ref{fig:library sig}, each gas' maximum absorption is at a different order of magnitude ranging from $10^{-6}$ to $10^{-2}$.
Since each gas was simulated to have a similar thermal contrast term, the choice of $n_c$ will largely determine the strength of the signal, and it will be different depending on the maximum absorption coefficient.
Using ACE detection at a threshold to produce a constant false alarm rate of $.5\%$, we found the $n_c$'s that resulted in an average true positive rate from a simulation of 100 plumes for each gas.

This allowed us to create several categorize of signal strength that can be compared across the different gases.
We chose to explore true positive rates ranging from 10\%, representative of a very weak signal, to 80\%, indicative of a very strong signal.
Simulating across 10 images, eight gases, and eight signal strength categories, we produced 640 plumes, allowing us to determine whether there are performance or hyperparameter trends for different gases or signal strengths.

\section{Results}
\label{sec:Results}

We are interested in two aspects of background estimation.
The first is how well a method estimates the true background radiance underneath a plume.
In theory, the better the background radiance estimate, the better the detection and identification performance.
We are also interested in how a method is able to change the NNI prediction confidence of the true gas.
The goal is to determine which methods will be able to increase the identification confidence, and lead to more accurate analyses and conclusions.

\subsection{Background Estimation}

\begin{figure}
    \centering
    \includegraphics[width=\linewidth]{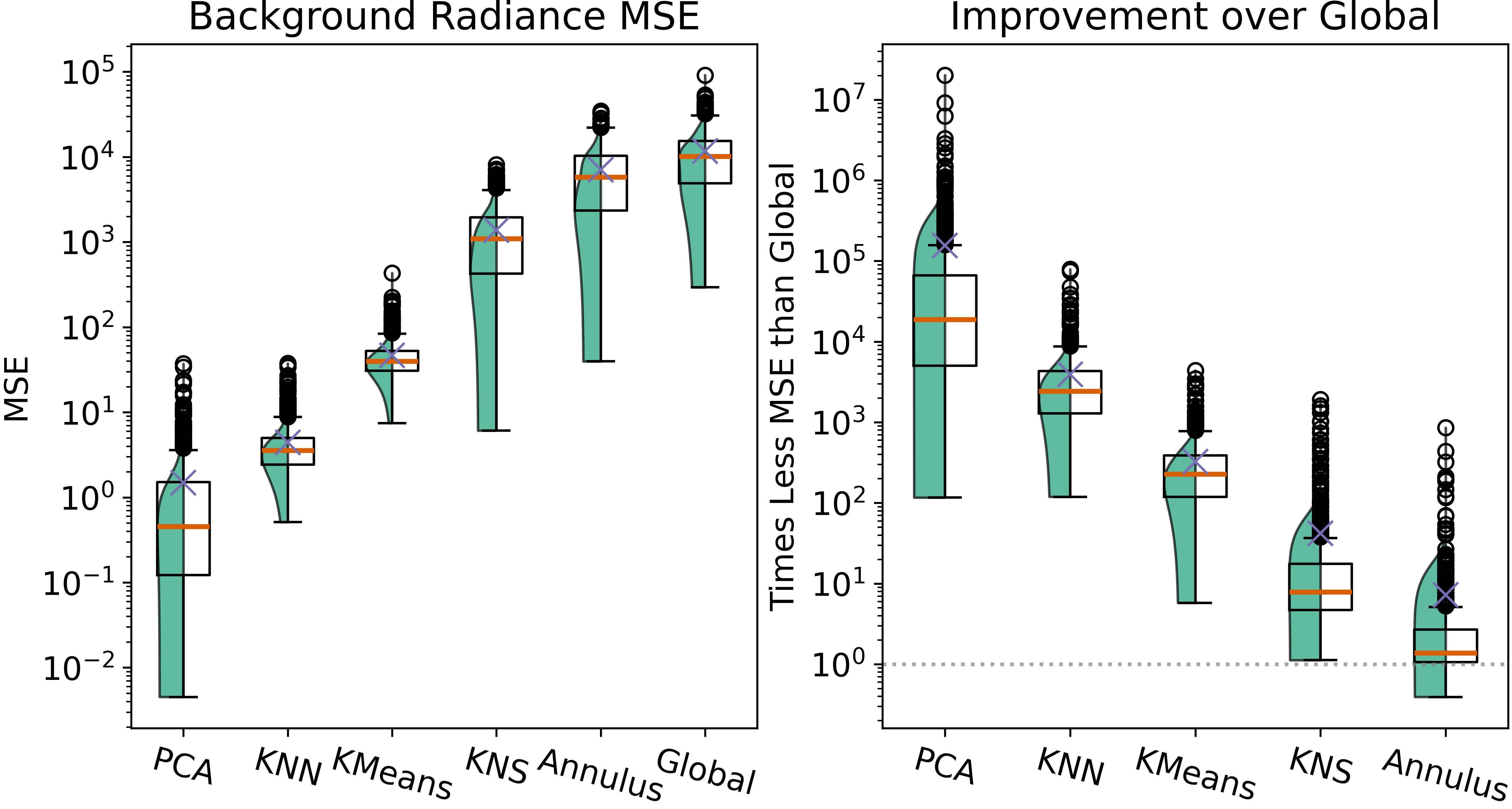}
    \caption{Background estimation MSE aggregated across all 640 simulated plumes for each method. Green shaded area are the kernel density estimates of the data. The orange lines indicate the median values and the purple ``x"s indicate the means. The dashed gray line at one indicates a method MSE equal to the corresponding Global MSE.}
    \label{fig:bg mse}
\end{figure}

The first task we investigate is in background radiance estimation.
For this, we measure the MSE between the true background radiance and the estimated background radiance for each pixel in a plume, with the goal to minimize the MSE. 
We first compare the raw MSE values for each method aggregated across all 640 images.
Figure \ref{fig:bg mse} shows the distribution of MSE values of each plume for each method, along with each method's respective MSE relative to Global MSE for the specific plume.
For example, for a specific plume, an improvement of two indicates the method has half the MSE compared to Global.

Overall, PCA has the best background estimation performance, followed closely by KNN, then KMeans, KNS, and Annulus.
The spatial methods are less performant than the non-spatial methods, though all methods generally perform better than Global.
In Figure \ref{fig:bg mse} each method has approximately an order of magnitude more MSE than the previous going from left to right.
KNN and KMeans have the least spread in MSE, meaning they have a fairly consistent background estimation performance.
This is in contrast to PCA, KNS, and Annulus that have larger variances, meaning they can have small MSE, but have a larger median MSE.

The right panel of Figure \ref{fig:bg mse} shows the ratio of the Global MSE and the method's MSE for each plume.
The trend matches what was seen previously, where PCA has the largest improvement over Global, with Annulus having the least improvement.
The dashed line indicates a method MSE that is equal to the corresponding global MSE.
Annulus is the only method that occasionally has a larger MSE than Global because there are some improvement values less than one.
PCA has the largest variance in improvement, ranging from as low as 116 all the way up to 2e7, with a median of around 18,800 times less MSE than Global.
Table \ref{tab:bg medians} shows the median MSE and improvement values for each method.

\begin{table}
    \centering
    \caption{Median background estimation MSE and improvement}
    \label{tab:bg medians}
    \begin{tabular}{r|r|r}
        Method       & MSE      & Improvement\\
        \hline
        \textbf{PCA} & 0.5      & 18,855.1\\
        \textit{KNN} & 3.6      & 2,420.6\\
        KMeans       & 39.6     & 227.8\\
        KNS          & 1,098.7  & 7.9\\
        Annulus      & 5,778.9  & 1.4\\
        Global       & 10,159.7 & \\
    \end{tabular}
\end{table}

\begin{figure}
    \centering
    \includegraphics[width=\linewidth]{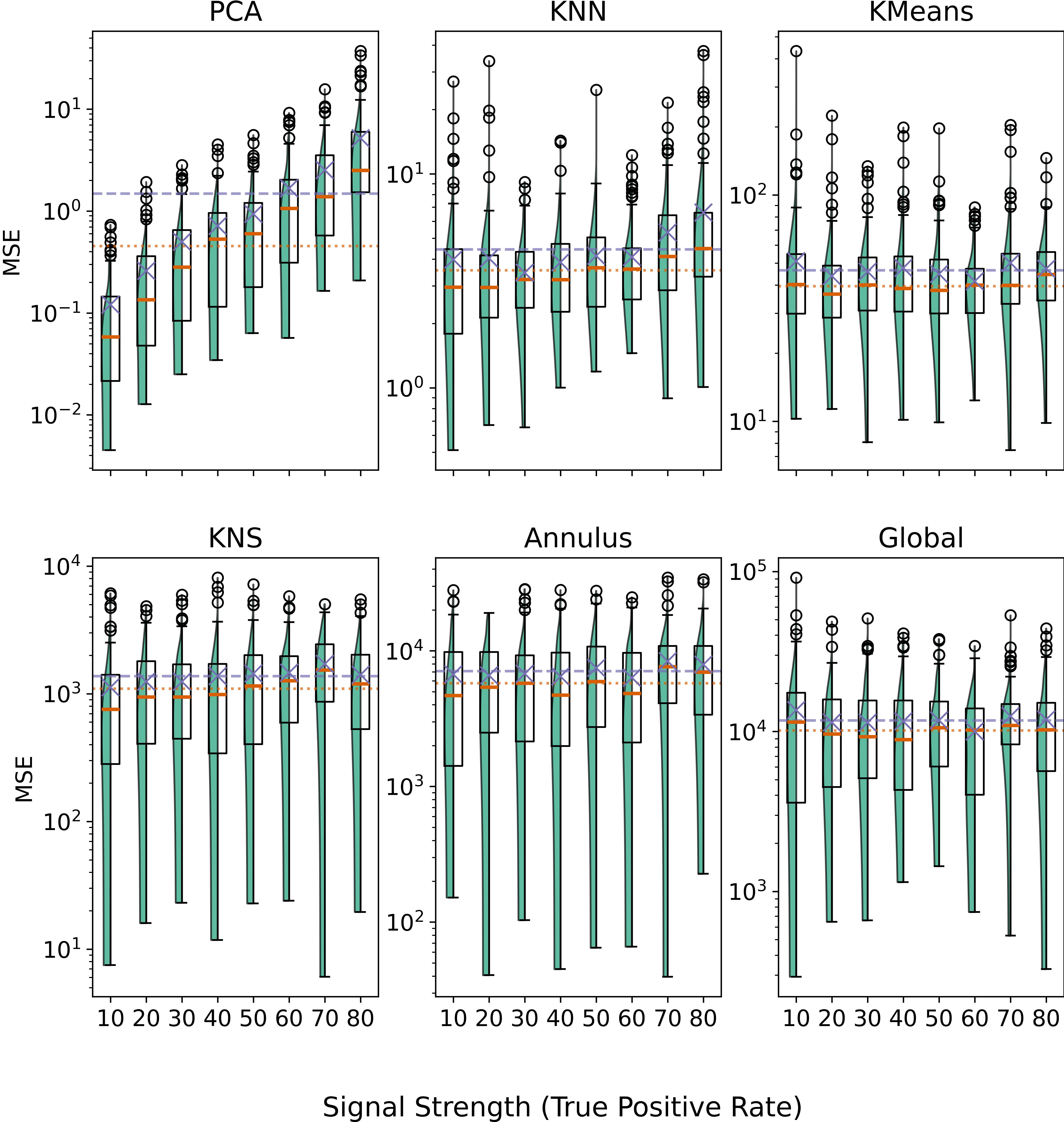}
    \caption{Plot of the distribution of MSE scores for each method when aggregated by signal strength. The horizontal dashed purple line is the average over all plumes, and the dotted orange line is the median over all plumes. The green areas are the density estimates of the distributions. Note the different scalings on the y-axes.}
    \label{fig:bg mse strength}
\end{figure}

We are also interested in the impact that signal strength has on the performance of each method.
Figure \ref{fig:bg mse strength} shows the MSE of each method when aggregated by signal strength.
Despite being the best performing method, PCA is most influenced by signal contamination having an exponential increase in MSE as signal strength increases.
KNN and KNS have increasing MSE trends, however, they are less prominent than that of PCA.
This suggests that these methods will benefit the most from having \textit{a priori} information, where the wavelengths that the gas has its strongest spectral features are ignored.

KMeans, Annulus, and Global have little, if any, trend with signal strength.
This suggests these methods may be more suited to the general case of no \textit{a priori} information.
However, compared to PCA and KNN, these methods do not perform as well in estimating the true background radiance values.

\subsection{Identification Confidence}

Next, we consider how different background estimates change our NNI confidence.
This value is found as the softmax of the final linear layer of the network, which results in a percentage value between 0$-$100\% representing how strongly the network assigns confidence to a given class.
We want to maximize this value, as it suggests the network is more confident in its prediction of the true gas.

\begin{figure}
    \centering
    \includegraphics[width=\linewidth]{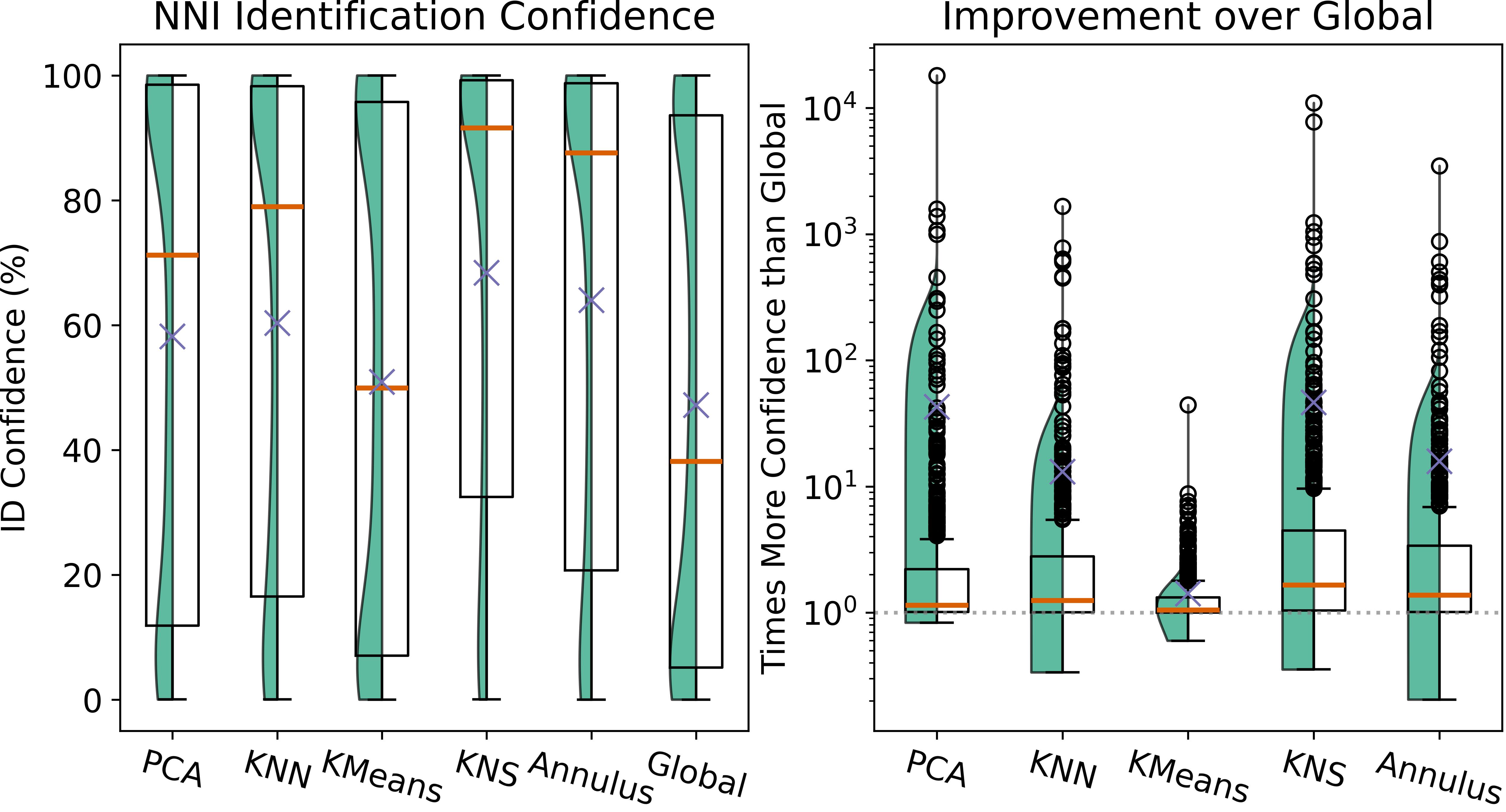}
    \caption{Left: Distribution of NNI confidence across all 640 plumes and each method. Right: distribution of how many times more confident the network is compared to Global. Orange lines are the median, purple ``x"s are the averages.}
    \label{fig:id conf}
\end{figure}

Similar to the background estimation task, NNI confidence is calculated for each plume and the distribution of these confidences is examined, presented in Figure \ref{fig:id conf}.
KNS results in the overall highest confidence, followed by Annulus, KNN, PCA, and lastly KMeans.
Again, each method generally offers higher confidence compared to Global, though the spatial methods outperform the non-spatial methods.
That is, the top two performing methods for identification are the bottom two performing methods in background estimation.

\begin{table}
    \centering
    \caption{Median identification confidence and improvement}
    \label{tab:id medians}
    \begin{tabular}{r|r|r}
        Method & ID Confidence & Improvement\\
        \hline
        PCA     & 71.2\% & 1.15\\
        KNN     & 79.0\% & 1.25\\
        KMeans  & 49.9\% & 1.05\\
        \textbf{KNS}     & 91.4\% & 1.65\\
        \textit{Annulus} & 87.6\% & 1.38\\
        Global  & 38.2\% & \\
    \end{tabular}
\end{table}

Regarding the improvement factor over Global, the values are much closer to unity than for MSE.
This is because NNI confidence is bounded above, resulting in improvement ratios being generally lower.
In the cases where improvement factors are in the range 10$-$100, Global confidence is exceptionally low.
Median performance improvement again shows that KNS and Annulus result in greater improvements.
It should be noted that the 25$^{th}$ percentile for each method is slightly above unity, indicating each method is expected to improve confidence over global about 75\% of the time.

\begin{figure}
    \centering
    \includegraphics[width=.5\linewidth]{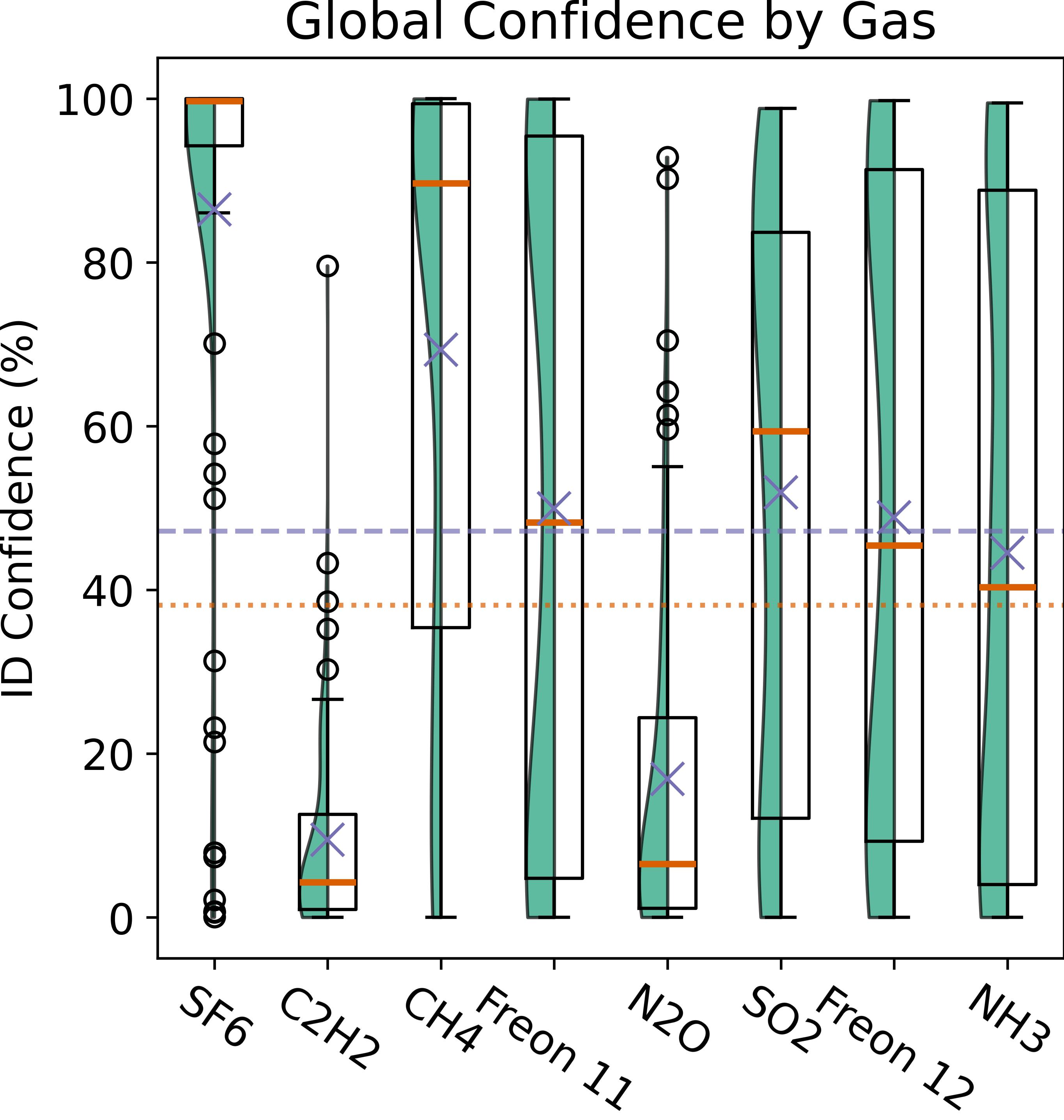}
    \caption{Distribution of NNI confidence by gas when using Global background estimation. The dashed purple line is the average confidence for all gases, and the dotted orange line is the median confidence for all gases.}
    \label{fig:id global gas}
\end{figure}

Our NNI generally does not have the same level of confidence for each gas.
Figure \ref{fig:id global gas} shows the distribution of confidences for each gas when using Global background estimation.
Our NNI has high confidence for SF$_6$ and CH$_4$ regardless of signal strength, whereas the NNI has low confidence for C$_2$H$_2$ and N$_2$O regardless of signal strength.
The remaining gases have near uniform confidence, meaning the model increases in confidence as signal strength increases.
It should be noted that this difference in confidence by gas is likely unique to our model specifically, and not necessarily the different spectral characteristics of the gases.

\begin{figure}
    \centering
    \includegraphics[width=\linewidth]{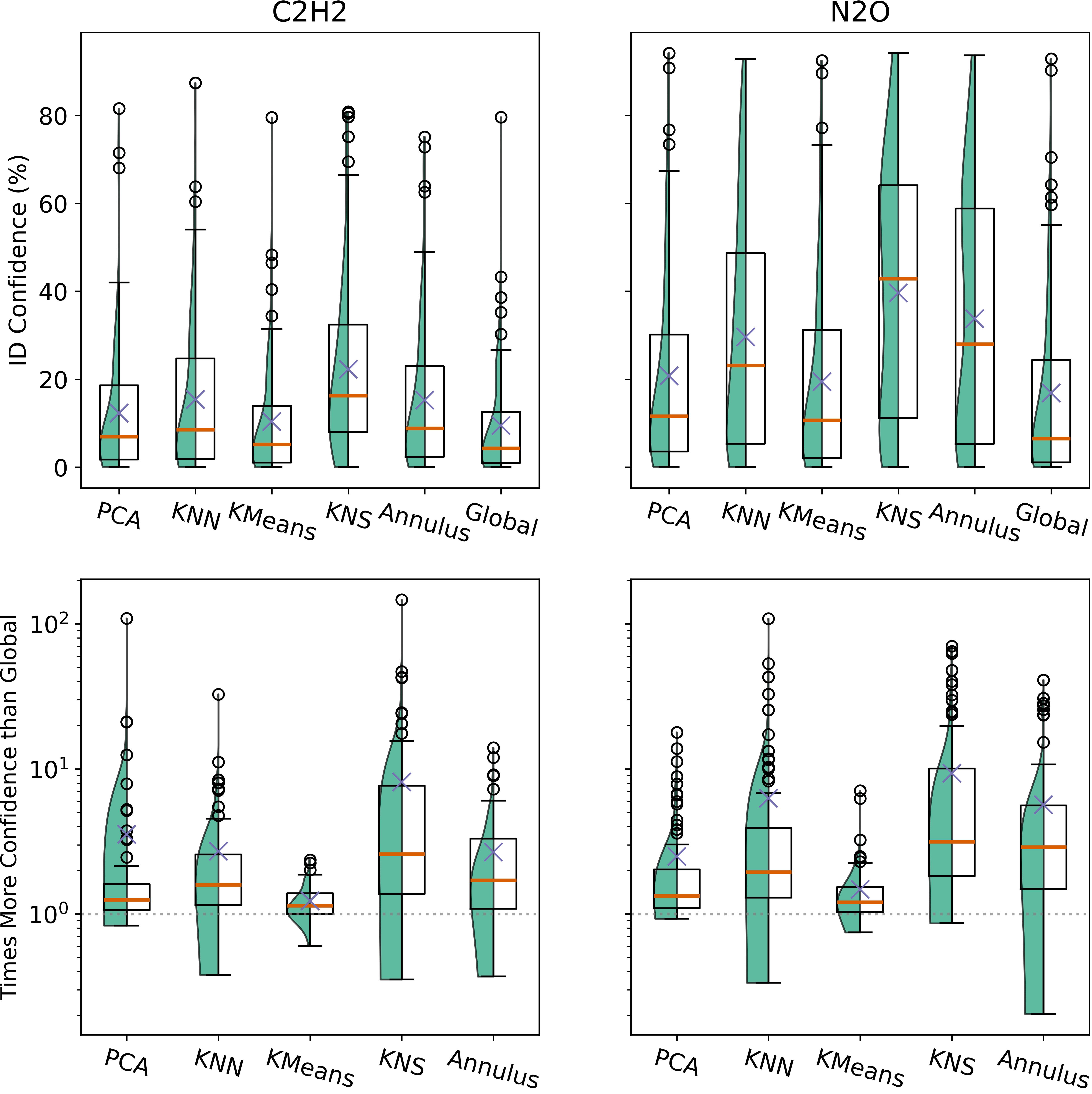}
    \caption{Identification confidence for hard to identify gases C$_2$H$_2$ and N$_2$O. The top row shows the raw confidence values, the bottom row shows the times improvement compared to Global.}
    \label{fig:id hardgas}
\end{figure}

Because the NNI has the most difficulty in identifying C$_2$H$_2$ and N$_2$O, it is helpful to understand how each method improves confidence for these gases specifically.
Figure \ref{fig:id hardgas} shows how each method performs when identifying C$_2$H$_2$ and N$_2$O.
The median Global identification confidence for each gas is 4.3\% and 6.5\% respectively.
Each method generally improves upon this, with KNS having median confidences of 16.3\% and 42.9\%.
In terms of the improvement factor ratio, KNS has a median of 2.6 and 3.1, followed closely by Annulus with 1.7 and 2.9.

These results demonstrate the benefits of using non-Global background estimation techniques.
Particularly, the spatial methods KNS and Annulus provide the largest benefits in identification performance.
These methods, in addition to raising general identification confidence, provide notable increases in confidence for gases which are difficult to identify.

\subsection{Hyperparameter Selection}
\label{sec:hyperparameter}

Each background estimation method requires a choice of one hyperparameter, with the exception of KNS which requires three.
Since each plume is different, the optimal choice of hyperparameter will likely be different for each plume.
In the aggregate of 640 plumes, there are common hyperparameter choices which can be used as default values, guiding hyperparameter selection for new plumes.

For each method and each plume, a grid search of hyperparameters was tested to determine which results in the best performance \cite{Bischl2021HyperparameterOF}.
For background estimation, the search finds the hyperparameter that produces the smallest MSE, while for identification confidence, the search finds the hyperparameter that produces the largest confidence.
We used the \texttt{Optuna} package in \texttt{Python} to facilitate the hyperparameter search \cite{Akiba2019OptunaAN}.
All of the previous figures illustrate results using the optimal hyperparameter for each plume.

For PCA, the number of components used was found within the range 1$-$127.
For KNN, the number of neighbors was found within the range 1$-$127.
For KMeans, the number of clusters was found within the range 2$-$128.
For Annulus, the number of dilations was found within the range 1$-$127.
For KNS, the number of minimum number of pixels found for values $2^n$ with $n\in[2,11]$.
The linkage function options are either single, complete, or average.
Lastly, there is a binary choice of whether or not to use BTS, where true is to use BTS, and false is to not use BTS.

Table \ref{tab:hyperparameter} shows the most common choices of hyperparameters, as well as the median value for each hyperparameter where applicable.
We find that for background estimation, PCA, KMeans, and KNS prefer many components, clusters, and pixels respectively.
In contrast, KNN and Annulus prefer few neighbors and dilations, respectively.
However, for identification every method prefers a lower hyperparameter value.
For KNS's two other hyperparameters, average linkage is preferred for background estimation while single linkage is preferred for identification confidence.
Similarly, using BTS is preferred for background estimation, but not preferred for identification.

\begin{table}
    \centering
    \caption{Common hyperparameter choices across all plumes}
    \label{tab:hyperparameter}
    \begin{tabular}{r|rr|rr}
         & \multicolumn{2}{c|}{Background} & \multicolumn{2}{c}{Identification}\\
        Method & Mode & Median & Mode & Median\\
        \hline
        PCA & 127 & 127 & 1 & 26\\
        KNN & 6 & 9 & 1 & 5\\
        KMeans & 128 & 124 & 2 & 77\\
        Annulus & 1 & 10 &  1 & 5\\
        \cdashline{1-5}
        \multirow{3}{*}{KNS} & 2048 & 256 & 4 & 16\\
        & Average & & Single\\
        & True & & False\\
    \end{tabular}
\end{table}

\begin{figure}
    \centering
    \includegraphics[width=\linewidth]{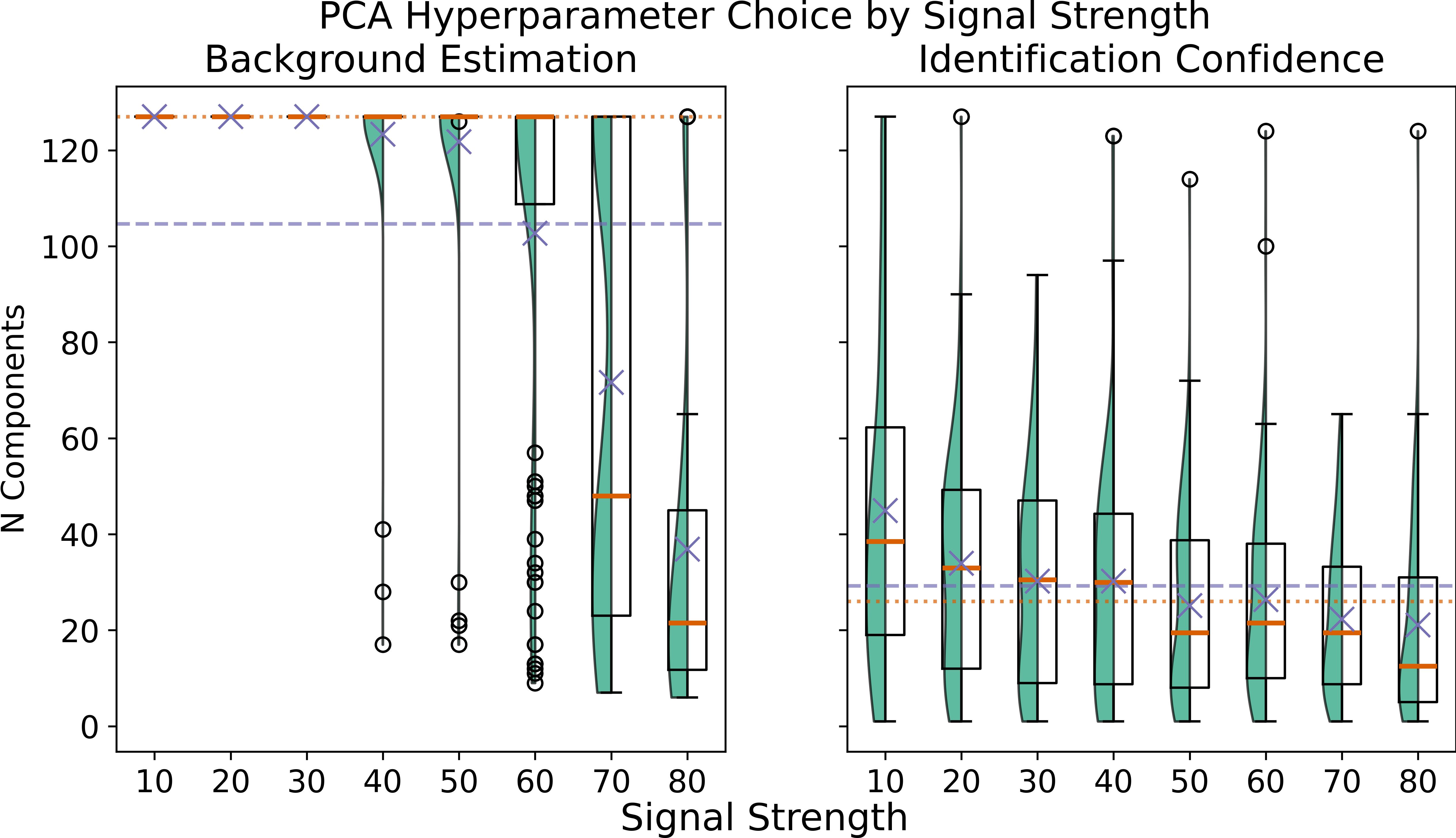}
    \caption{Optimal choice of the number of components for PCA for both background estimation and identification confidence.}
    \label{fig:pca params}
\end{figure}

It was found previously that PCA produces the best background radiance estimates, however, MSE increases as signal strength increases.
There is also a hyperparameter trend for PCA with signal strength as shown in Figure \ref{fig:pca params}.
For background estimation, 127 components is chosen for all plumes with strengths in the range 10$-$30.
In the signal strength range 70$-$80, there is a preference for fewer components, specifically medians of 48 and 22 components respectively.
This trend is also present in the identification confidence task, however, weak plumes prefer around 40 components, while strong plumes prefer around 10 components.

KNS is found to generally produce background radiance estimates slightly better than Global and to notably increase identification confidence.
The optimal hyperparameters for background estimation and identification confidence were found to be different, particularly in whether or not to use BTS.
Table \ref{tab:kns params} summarizes how many times each hyperparameter option was found to be the best performing choice.
We find that for background estimation, BTS is only slightly preferred being used 51.4\% of the time, while for identification confidence BTS is only preferred 35.9\% of the time.
Similarly, average linkage is preferred for background estimation, while single linkage is only slightly preferred for identification confidence.
This suggests that for identification confidence, BTS is not optimal, which will reduce computation time, and the choice of linkage function is not as important.

\begin{table}
    \centering
    \caption{KNS Hyperparameter Choices}
    \label{tab:kns params}
    \begin{tabular}{rl|r|r}
        \multicolumn{2}{c|}{Hyperparameter} & Background & Identification\\
        \hline
        \multirow{3}{*}{Linkage:}    & Single   & 90 (14.1\%)  & \textbf{268 (41.9\%)}\\
            & Complete & 111 (17.3\%) & 211 (33.0\%)\\
            & Average  & \textbf{439 (68.6\%)} & 161 (25.1\%)\\
        \cdashline{1-4}
        \multirow{2}{*}{Use BTS?:} & True & \textbf{329 (51.4\%)} & 230 (35.9\%)\\
                & False & 311 (48.6\%) & \textbf{410 (64.1\%)}
    \end{tabular}
\end{table}

This investigation highlights how sensitive each method is to its choice of hyperparameter.
It would be preferred that a method not be sensitive to its choice of hyperparameter, as this would make the method easier to apply to many plumes.
Taking the standard deviation (STD) of the response values from our hyperparameter search can give a measure of how much the response varies with different choices of hyperparameters.
For example, considering background estimation, if a method has a small STD of the MSE values from all choices of hyperparameters, this suggests the method is not sensitive to the specific choice of hyperparameter, as each hyperparameter produces a similar MSE.

\begin{figure}
    \centering
    \includegraphics[width=\linewidth]{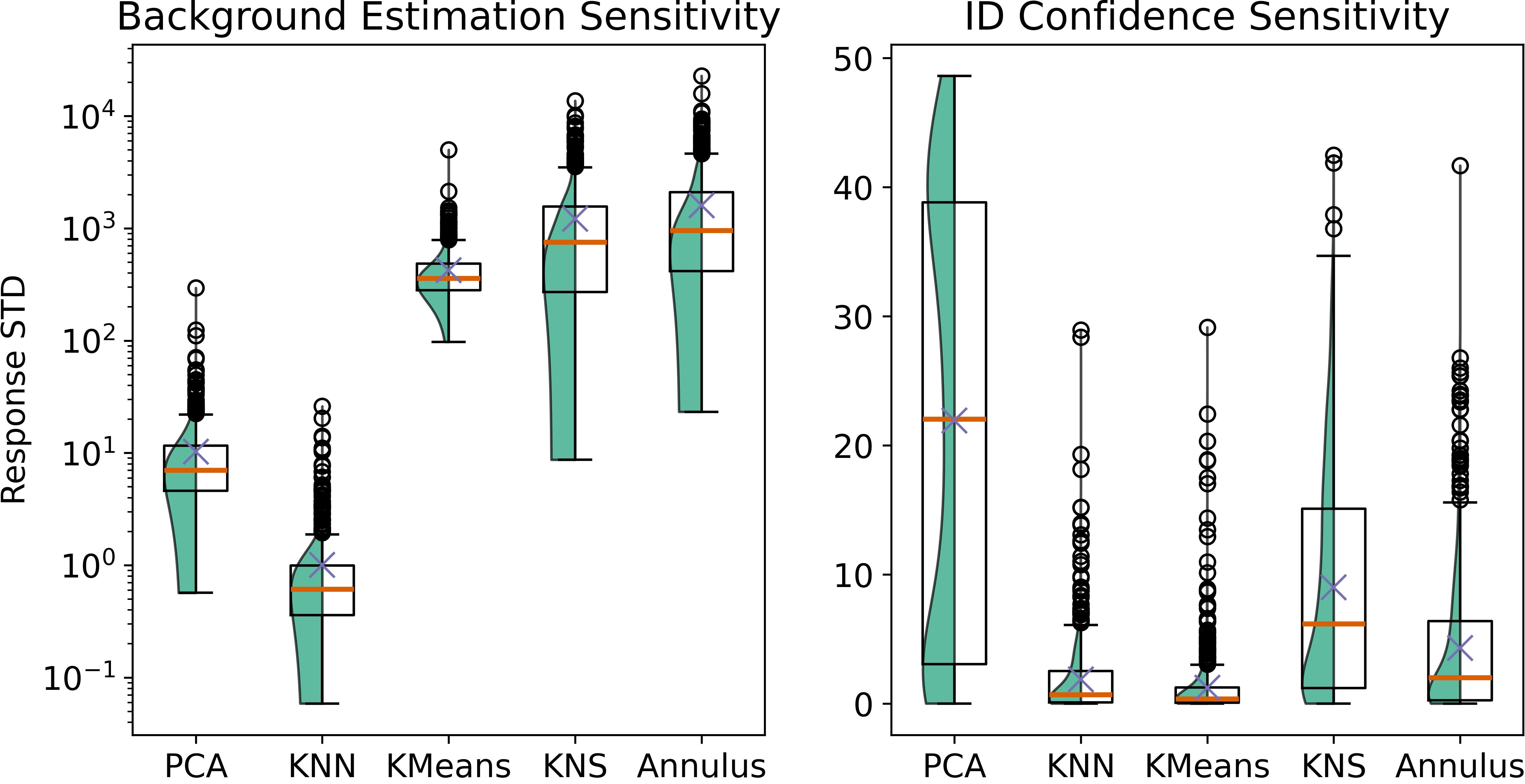}
    \caption{Hyperparameter sensitivity for each method. The right plot shows the standard deviation of the MSE values from the grid search. The left plot shows the standard deviation of the identification confidence values from the grid search.}
    \label{fig:param sens}
\end{figure}

Figure \ref{fig:param sens} shows the distribution of how sensitive each method is to the choice of hyperparameters for each task.
KNN and PCA have the least sensitivity to their respective hyperparmeters for background estimation, while KMeans and KNN have the least sensitivity for identification confidence.
KNS and Annulus have generally high sensitivity, but also have a wide range of sensitivity values, meaning these methods are occasionally insensitive to some plumes.
Looking at identification confidence, PCA has the largest sensitivity, meaning the choice of number of components has a large impact on how beneficial PCA is to identification confidence.
The best performing methods KNS and Annulus have the next highest sensitivities.
This means, to get the best improvement in identification confidence, some hyperparameter tuning may be required.
However, their overall sensitivity values are low, suggesting these methods may still provide reasonable performance using default hyperparameters.

Because KNN has low sensitivity for both background estimation and identification confidence, we consider it a good ``out of the box" method.
This is because the choice of the hyperparameter $k$ does not drastically change background MSE or identification confidence, and based on Table \ref{tab:hyperparameter} a good default hyperparameter for both tasks could be the same, perhaps $k=8$.
Furthermore, KNN has the second best background estimation performance, and third largest improvement for identification confidence.

\section{Conclusions}
\label{sec:Conclusions}

This paper has three contributions to the field of gas plume identification.
First is the testing of non-global background estimation strategies for identification, particularly when trying to use a neural network to identify a gas from a large library of candidates. 
Second is the extension and comparison of new and existing background estimation methods to identification, including a comparison between spatial and non-spatial methods.
Third, we simulated a large number of gas plumes to analyze trends and make practical conclusions for the applicability of each method.

After simulating 640 plumes of varying gases and signal strengths to evaluate how each method performs across a range of different plume morphologies, we find that PCA produces the best background radiance estimates, having a median of 18,000 times less MSE compared to Global.
PCA, KNN, and KNS were found to have increased MSE as signal strength increased, suggesting a benefit from the use of \textit{a priori} information when available.
In addition to a set of common hyperparameters for each method, we found that PCA and KNN are relatively insensitive to the choice of hyperparameter.

However, for identification confidence, KNS was found to improve neural network confidence of the true gas the most, going from a 38.2\% median confidence using Global estimation, to 91.4\% confidence.
Furthermore, KNS is particularly helpful for hard to identify gases.
Every method was found to prefer small values of hyperparameters.
The best performing methods KNS and Annulus were found to have relatively high hyperparameter sensitivity, suggesting that in practice some hyperparameter tuning is likely required for optimal performance.

We found that KNN is a good ``out of the box" option for both background estimation and identification confidence.
This is because it has low hyperparameter sensitivity, a consistent default hyperparameter choice, and it has the second best background estimation performance and third best identification confidence.

There are several directions for future work.
Investigating the inclusion of a local covariance estimate in addition to a local mean estimate is needed, or even an alternative to the whitening paradigm.
Additionally, it is curious that the best methods for identification confidence were consistently the least performant background estimation methods.
Further investigation is needed in how to incorporate PCA or KNN into the training of our NNI, such that the best background estimation methods lead to the highest identification confidence. 
Lastly, we conducted these experiments on simulated plumes.
More research is needed on real world plumes to determine the practical applicability of these algorithms, with ground truth spectral references.





\section*{Acknowledgment}

Thank you to Dr. Eric Flynn and Dr. James Theiler for their insights and conversations about background estimation.
The authors acknowledge the Aerospace Corporation for collecting and providing the historical airborne LWIR data from the Los Angeles basin area. This work is partially funded by NNSA Contract No. 89233218CNA000001, and Los Alamos National Laboratory subcontract No. C3486. LA-UR-24-31408 Ver. 2.

\ifCLASSOPTIONcaptionsoff
  \newpage
\fi



\bibliographystyle{IEEEtran}
\bibliography{IEEEabrv,bib}

\begin{thebibliography}{10}
\providecommand{\url}[1]{#1}
\csname url@samestyle\endcsname
\providecommand{\newblock}{\relax}
\providecommand{\bibinfo}[2]{#2}
\providecommand{\BIBentrySTDinterwordspacing}{\spaceskip=0pt\relax}
\providecommand{\BIBentryALTinterwordstretchfactor}{4}
\providecommand{\BIBentryALTinterwordspacing}{\spaceskip=\fontdimen2\font plus
\BIBentryALTinterwordstretchfactor\fontdimen3\font minus \fontdimen4\font\relax}
\providecommand{\BIBforeignlanguage}[2]{{%
\expandafter\ifx\csname l@#1\endcsname\relax
\typeout{** WARNING: IEEEtran.bst: No hyphenation pattern has been}%
\typeout{** loaded for the language `#1'. Using the pattern for}%
\typeout{** the default language instead.}%
\else
\language=\csname l@#1\endcsname
\fi
#2}}
\providecommand{\BIBdecl}{\relax}
\BIBdecl

\bibitem{Stuart2019HyperspectralII}
\BIBentryALTinterwordspacing
M.~B. Stuart, A.~J.~S. McGonigle, and J.~R. Willmott, ``Hyperspectral imaging in environmental monitoring: A review of recent developments and technological advances in compact field deployable systems,'' \emph{Sensors (Basel, Switzerland)}, vol.~19, 2019. [Online]. Available: \url{https://api.semanticscholar.org/CorpusID:198192705}
\BIBentrySTDinterwordspacing

\bibitem{Pallua2021NewPO}
\BIBentryALTinterwordspacing
J.~D. Pallua, A.~Brunner, B.~Zelger, C.~W. Huck, M.~Schirmer, J.~Laimer, D.~Putzer, M.~Thaler, and B.~Zelger, ``New perspectives of hyperspectral imaging for clinical research,'' \emph{NIR News}, vol.~32, pp. 5 -- 13, 2021. [Online]. Available: \url{https://api.semanticscholar.org/CorpusID:235760916}
\BIBentrySTDinterwordspacing

\bibitem{Yuen2010AnIT}
\BIBentryALTinterwordspacing
P.~W.~T. Yuen and M.~A. Richardson, ``An introduction to hyperspectral imaging and its application for security, surveillance and target acquisition,'' \emph{The Imaging Science Journal}, vol.~58, pp. 241 -- 253, 2010. [Online]. Available: \url{https://api.semanticscholar.org/CorpusID:34821906}
\BIBentrySTDinterwordspacing

\bibitem{Khan2018ModernTI}
\BIBentryALTinterwordspacing
M.~J. Khan, H.~S. Khan, A.~Yousaf, K.~Khurshid, and A.~Abbas, ``Modern trends in hyperspectral image analysis: A review,'' \emph{IEEE Access}, vol.~6, pp. 14\,118--14\,129, 2018. [Online]. Available: \url{https://api.semanticscholar.org/CorpusID:4624178}
\BIBentrySTDinterwordspacing

\bibitem{Hulley2016HighSR}
\BIBentryALTinterwordspacing
G.~C. Hulley, R.~M. Duren, F.~M. Hopkins, S.~J. Hook, N.~Vance, P.~C. Guillevic, W.~R. Johnson, B.~T. Eng, J.~M. Mihaly, V.~M. Jovanovic, S.~L. Chazanoff, Z.~K. Staniszewski, L.~Kuai, J.~R. Worden, C.~Frankenberg, G.~Rivera, A.~D. Aubrey, C.~E. Miller, N.~K. Malakar, J.~M.~S. Tom{\'a}s, and K.~T. Holmes, ``High spatial resolution imaging of methane and other trace gases with the airborne hyperspectral thermal emission spectrometer (hytes),'' \emph{Atmospheric Measurement Techniques}, vol.~9, pp. 2393--2408, 2016. [Online]. Available: \url{https://api.semanticscholar.org/CorpusID:54893595}
\BIBentrySTDinterwordspacing

\bibitem{Burr2006OverviewOP}
\BIBentryALTinterwordspacing
T.~L. Burr and N.~W. Hengartner, ``Overview of physical models and statistical approaches for weak gaseous plume detection using passive infrared hyperspectral imagery,'' \emph{Sensors}, vol.~6, pp. 1721--1750, 2006. [Online]. Available: \url{https://api.semanticscholar.org/CorpusID:6554182}
\BIBentrySTDinterwordspacing

\bibitem{Manolakis2014DetectionAI}
\BIBentryALTinterwordspacing
D.~G. Manolakis, E.~Truslow, M.~L. Pieper, T.~W. Cooley, and M.~Brueggeman, ``Detection algorithms in hyperspectral imaging systems: An overview of practical algorithms,'' \emph{IEEE Signal Processing Magazine}, vol.~31, pp. 24--33, 2014. [Online]. Available: \url{https://api.semanticscholar.org/CorpusID:18877994}
\BIBentrySTDinterwordspacing

\bibitem{Manolakis2002DetectionAF}
\BIBentryALTinterwordspacing
D.~G. Manolakis and G.~A. Shaw, ``Detection algorithms for hyperspectral imaging applications,'' \emph{IEEE Signal Process. Mag.}, vol.~19, pp. 29--43, 2002. [Online]. Available: \url{https://api.semanticscholar.org/CorpusID:62194350}
\BIBentrySTDinterwordspacing

\bibitem{Truslow2016PerformanceMF}
\BIBentryALTinterwordspacing
E.~Truslow, S.~E. Golowich, D.~G. Manolakis, and V.~K. Ingle, ``Performance metrics for the evaluation of hyperspectral chemical identification systems,'' \emph{Optical Engineering}, vol.~55, 2016. [Online]. Available: \url{https://api.semanticscholar.org/CorpusID:53578703}
\BIBentrySTDinterwordspacing

\bibitem{Burr2002ChemicalIU}
\BIBentryALTinterwordspacing
T.~L. Burr, H.~Fry, B.~McVey, and E.~Sander, ``Chemical identification using bayesian model selection,'' 2002. [Online]. Available: \url{https://api.semanticscholar.org/CorpusID:17843691}
\BIBentrySTDinterwordspacing

\bibitem{Hoeting1999BayesianMA}
\BIBentryALTinterwordspacing
J.~A. Hoeting, D.~Madigan, A.~E. Raftery, and C.~Volinsky, ``Bayesian model averaging: A tutorial,'' 1999. [Online]. Available: \url{https://api.semanticscholar.org/CorpusID:60698623}
\BIBentrySTDinterwordspacing

\bibitem{Pogorzala2005GasPS}
\BIBentryALTinterwordspacing
D.~R. Pogorzala, D.~W. Messinger, C.~Salvaggio, and J.~R. Schott, ``Gas plume species identification in airborne lwir imagery using constrained stepwise regression analyses,'' 2005. [Online]. Available: \url{https://api.semanticscholar.org/CorpusID:11211080}
\BIBentrySTDinterwordspacing

\bibitem{Li2019DeepLF}
\BIBentryALTinterwordspacing
S.~Li, W.~Song, L.~Fang, Y.~Chen, P.~Ghamisi, and J.~A. Benediktsson, ``Deep learning for hyperspectral image classification: An overview,'' \emph{IEEE Transactions on Geoscience and Remote Sensing}, vol.~57, pp. 6690--6709, 2019. [Online]. Available: \url{https://api.semanticscholar.org/CorpusID:150101748}
\BIBentrySTDinterwordspacing

\bibitem{Klein2023HyperspectralTI}
\BIBentryALTinterwordspacing
N.~Klein, A.~Carr, Z.~Hampel-Arias, A.~Zastrow, A.~Ziemann, and E.~Flynn, ``Hyperspectral target identification using physics-guided neural networks with explainability and feature attribution,'' \emph{IGARSS 2023 - 2023 IEEE International Geoscience and Remote Sensing Symposium}, pp. 946--949, 2023. [Online]. Available: \url{https://api.semanticscholar.org/CorpusID:264356064}
\BIBentrySTDinterwordspacing

\bibitem{Idoughi2016BackgroundRE}
\BIBentryALTinterwordspacing
R.~Idoughi, T.~H.~G. Vidal, P.-Y. Foucher, M.~A. Gagnon, and X.~Briottet, ``Background radiance estimation for gas plume quantification for airborne hyperspectral thermal imaging,'' \emph{Spectroscopy}, vol. 2016, pp. 1--17, 2016. [Online]. Available: \url{https://api.semanticscholar.org/CorpusID:53558315}
\BIBentrySTDinterwordspacing

\bibitem{Hayden1997RemoteTG}
\BIBentryALTinterwordspacing
A.~F. Hayden and R.~J. Noll, ``Remote trace gas quantification using thermal ir spectroscopy and digital filtering based on principal components of background scene clutter,'' in \emph{Defense, Security, and Sensing}, 1997. [Online]. Available: \url{https://api.semanticscholar.org/CorpusID:128877914}
\BIBentrySTDinterwordspacing

\bibitem{Manolakis2019LongwaveIHPPC}
\BIBentryALTinterwordspacing
D.~G. Manolakis, M.~L. Pieper, E.~Truslow, R.~B. Lockwood, A.~Weisner, J.~Jacobson, and T.~W. Cooley, ``Longwave infrared hyperspectral imaging: Principles, progress, and challenges,'' \emph{IEEE Geoscience and Remote Sensing Magazine}, vol.~7, pp. 72--100, 2019. [Online]. Available: \url{https://api.semanticscholar.org/CorpusID:195222649}
\BIBentrySTDinterwordspacing

\bibitem{Manolakis2014LongWaveIH}
\BIBentryALTinterwordspacing
D.~G. Manolakis, S.~E. Golowich, and R.~S. DiPietro, ``Long-wave infrared hyperspectral remote sensing of chemical clouds: A focus on signal processing approaches,'' \emph{IEEE Signal Processing Magazine}, vol.~31, pp. 120--141, 2014. [Online]. Available: \url{https://api.semanticscholar.org/CorpusID:16761725}
\BIBentrySTDinterwordspacing

\bibitem{Buckland2017TrackingAQ}
\BIBentryALTinterwordspacing
K.~N. Buckland, S.~J. Young, E.~R. Keim, B.~R. Johnson, P.~D. Johnson, and D.~M. Tratt, ``Tracking and quantification of gaseous chemical plumes from anthropogenic emission sources within the los angeles basin,'' \emph{Remote Sensing of Environment}, vol. 201, pp. 275--296, 2017. [Online]. Available: \url{https://api.semanticscholar.org/CorpusID:134128864}
\BIBentrySTDinterwordspacing

\bibitem{Nasrabadi2014HyperspectralTD}
\BIBentryALTinterwordspacing
N.~M. Nasrabadi, ``Hyperspectral target detection : An overview of current and future challenges,'' \emph{IEEE Signal Processing Magazine}, vol.~31, pp. 34--44, 2014. [Online]. Available: \url{https://api.semanticscholar.org/CorpusID:12240632}
\BIBentrySTDinterwordspacing

\bibitem{Matteoli2014AnOO}
\BIBentryALTinterwordspacing
S.~Matteoli, M.~Diani, and J.~Theiler, ``An overview of background modeling for detection of targets and anomalies in hyperspectral remotely sensed imagery,'' \emph{IEEE Journal of Selected Topics in Applied Earth Observations and Remote Sensing}, vol.~7, pp. 2317--2336, 2014. [Online]. Available: \url{https://api.semanticscholar.org/CorpusID:18353243}
\BIBentrySTDinterwordspacing

\bibitem{Manolakis2009IsTA}
\BIBentryALTinterwordspacing
D.~G. Manolakis, R.~B. Lockwood, T.~W. Cooley, and J.~Jacobson, ``Is there a best hyperspectral detection algorithm?'' in \emph{Defense + Commercial Sensing}, 2009. [Online]. Available: \url{https://api.semanticscholar.org/CorpusID:26987112}
\BIBentrySTDinterwordspacing

\bibitem{Carlotto2005ACA}
\BIBentryALTinterwordspacing
M.~J. Carlotto, ``A cluster-based approach for detecting man-made objects and changes in imagery,'' \emph{IEEE Transactions on Geoscience and Remote Sensing}, vol.~43, pp. 374--387, 2005. [Online]. Available: \url{https://api.semanticscholar.org/CorpusID:11003912}
\BIBentrySTDinterwordspacing

\bibitem{Marden2004UsingEC}
\BIBentryALTinterwordspacing
D.~Marden and D.~G. Manolakis, ``Using elliptically contoured distributions to model hyperspectral imaging data and generate statistically similar synthetic data,'' in \emph{SPIE Defense + Commercial Sensing}, 2004. [Online]. Available: \url{https://api.semanticscholar.org/CorpusID:122244177}
\BIBentrySTDinterwordspacing

\bibitem{Matteoli2013ModelsAM}
\BIBentryALTinterwordspacing
S.~Matteoli, T.~Veracini, M.~Diani, and G.~Corsini, ``Models and methods for automated background density estimation in hyperspectral anomaly detection,'' \emph{IEEE Transactions on Geoscience and Remote Sensing}, vol.~51, pp. 2837--2852, 2013. [Online]. Available: \url{https://api.semanticscholar.org/CorpusID:626234}
\BIBentrySTDinterwordspacing

\bibitem{Jarman2023EnsembleSF}
\BIBentryALTinterwordspacing
S.~Jarman, T.~Carr, Z.~Hampel-Arias, E.~Flynn, and K.~R. Moon, ``Ensemble segmentation for improved background estimation and gas plume identification in hyperspectral images,'' in \emph{Optical Engineering + Applications}, 2023. [Online]. Available: \url{https://api.semanticscholar.org/CorpusID:263689547}
\BIBentrySTDinterwordspacing

\bibitem{Hartigan1979AKC}
\BIBentryALTinterwordspacing
J.~A. Hartigan and M.~A. Wong, ``A k-means clustering algorithm,'' 1979. [Online]. Available: \url{https://api.semanticscholar.org/CorpusID:53880671}
\BIBentrySTDinterwordspacing

\bibitem{Arthur2007kmeansTA}
\BIBentryALTinterwordspacing
D.~Arthur and S.~Vassilvitskii, ``k-means++: the advantages of careful seeding,'' in \emph{ACM-SIAM Symposium on Discrete Algorithms}, 2007. [Online]. Available: \url{https://api.semanticscholar.org/CorpusID:1782131}
\BIBentrySTDinterwordspacing

\bibitem{Funk2001ClusteringTI}
\BIBentryALTinterwordspacing
C.~Funk, J.~Theiler, D.~A. Roberts, and C.~Borel-Donohue, ``Clustering to improve matched filter detection of weak gas plumes in hyperspectral thermal imagery,'' \emph{IEEE Trans. Geosci. Remote. Sens.}, vol.~39, pp. 1410--1420, 2001. [Online]. Available: \url{https://api.semanticscholar.org/CorpusID:722009}
\BIBentrySTDinterwordspacing

\bibitem{Jolliffe2002PrincipalCA}
\BIBentryALTinterwordspacing
I.~T. Jolliffe, ``Principal component analysis,'' 2002. [Online]. Available: \url{https://api.semanticscholar.org/CorpusID:265962117}
\BIBentrySTDinterwordspacing

\bibitem{Manolakis2001HyperspectralST}
\BIBentryALTinterwordspacing
D.~G. Manolakis, C.~Siracusa, and G.~A. Shaw, ``Hyperspectral subpixel target detection using the linear mixing model,'' \emph{IEEE Trans. Geosci. Remote. Sens.}, vol.~39, pp. 1392--1409, 2001. [Online]. Available: \url{https://api.semanticscholar.org/CorpusID:43627459}
\BIBentrySTDinterwordspacing

\bibitem{Hayden1996DeterminationOT}
\BIBentryALTinterwordspacing
A.~F. Hayden, E.~R. Niple, and B.~E. Boyce, ``Determination of trace-gas amounts in plumes by the use of orthogonal digital filtering of thermal-emission spectra.'' \emph{Applied optics}, vol. 35 16, pp. 2802--9, 1996. [Online]. Available: \url{https://api.semanticscholar.org/CorpusID:985274}
\BIBentrySTDinterwordspacing

\bibitem{Bao2016IntroductionTM}
\BIBentryALTinterwordspacing
W.~Bao, ``Introduction to machine learning: k-nearest neighbors.'' \emph{Annals of translational medicine}, vol. 4 11, p. 218, 2016. [Online]. Available: \url{https://api.semanticscholar.org/CorpusID:42966081}
\BIBentrySTDinterwordspacing

\bibitem{Ma2010LocalML}
\BIBentryALTinterwordspacing
L.~Ma, M.~M. Crawford, and J.~Tian, ``Local manifold learning-based \$k\$ -nearest-neighbor for hyperspectral image classification,'' \emph{IEEE Transactions on Geoscience and Remote Sensing}, vol.~48, pp. 4099--4109, 2010. [Online]. Available: \url{https://api.semanticscholar.org/CorpusID:29808774}
\BIBentrySTDinterwordspacing

\bibitem{Huang2015SpectralSpatialHI}
\BIBentryALTinterwordspacing
K.~Huang, S.~Li, X.~Kang, and L.~Fang, ``Spectral–spatial hyperspectral image classification based on knn,'' \emph{Sensing and Imaging}, vol.~17, 2015. [Online]. Available: \url{https://api.semanticscholar.org/CorpusID:123815316}
\BIBentrySTDinterwordspacing

\bibitem{Cui2015ClassDependentSR}
\BIBentryALTinterwordspacing
M.~Cui and S.~Prasad, ``Class-dependent sparse representation classifier for robust hyperspectral image classification,'' \emph{IEEE Transactions on Geoscience and Remote Sensing}, vol.~53, pp. 2683--2695, 2015. [Online]. Available: \url{https://api.semanticscholar.org/CorpusID:5214904}
\BIBentrySTDinterwordspacing

\bibitem{Guo2017KNearestNC}
\BIBentryALTinterwordspacing
Y.~Guo, S.~Han, Y.~Li, C.~Zhang, and Y.~Bai, ``K-nearest neighbor combined with guided filter for hyperspectral image classification,'' in \emph{International Conference on Identification, Information, and Knowledge in the Internet of Things}, 2017. [Online]. Available: \url{https://api.semanticscholar.org/CorpusID:51868811}
\BIBentrySTDinterwordspacing

\bibitem{Matteoli2014BackgroundDN}
\BIBentryALTinterwordspacing
S.~Matteoli, T.~Veracini, M.~Diani, and G.~Corsini, ``Background density nonparametric estimation with data-adaptive bandwidths for the detection of anomalies in multi-hyperspectral imagery,'' \emph{IEEE Geoscience and Remote Sensing Letters}, vol.~11, pp. 163--167, 2014. [Online]. Available: \url{https://api.semanticscholar.org/CorpusID:23215874}
\BIBentrySTDinterwordspacing

\bibitem{Matteoli2014ALA}
\BIBentryALTinterwordspacing
------, ``A locally adaptive background density estimator: An evolution for rx-based anomaly detectors,'' \emph{IEEE Geoscience and Remote Sensing Letters}, vol.~11, pp. 323--327, 2014. [Online]. Available: \url{https://api.semanticscholar.org/CorpusID:17278403}
\BIBentrySTDinterwordspacing

\bibitem{Acito2005AdaptiveDA}
\BIBentryALTinterwordspacing
N.~Acito, G.~Corsini, and M.~Diani, ``Adaptive detection algorithm for full pixel targets in hyperspectral images,'' 2005. [Online]. Available: \url{https://api.semanticscholar.org/CorpusID:140141650}
\BIBentrySTDinterwordspacing

\bibitem{Matteoli2010ATO}
\BIBentryALTinterwordspacing
S.~Matteoli, M.~Diani, and G.~Corsini, ``A tutorial overview of anomaly detection in hyperspectral images,'' \emph{IEEE Aerospace and Electronic Systems Magazine}, vol.~25, pp. 5--28, 2010. [Online]. Available: \url{https://api.semanticscholar.org/CorpusID:36740158}
\BIBentrySTDinterwordspacing

\bibitem{Grewal2022HyperspectralIS}
\BIBentryALTinterwordspacing
R.~Grewal, S.~S. Kasana, and G.~Kasana, ``Hyperspectral image segmentation: a comprehensive survey,'' \emph{Multimedia Tools and Applications}, vol.~82, pp. 20\,819 -- 20\,872, 2022. [Online]. Available: \url{https://api.semanticscholar.org/CorpusID:253231018}
\BIBentrySTDinterwordspacing

\bibitem{Tarabalka2010SegmentationAC}
\BIBentryALTinterwordspacing
Y.~Tarabalka, J.~Chanussot, and J.~A. Benediktsson, ``Segmentation and classification of hyperspectral images using watershed transformation,'' \emph{Pattern Recognit.}, vol.~43, pp. 2367--2379, 2010. [Online]. Available: \url{https://api.semanticscholar.org/CorpusID:26615105}
\BIBentrySTDinterwordspacing

\bibitem{Ding2002ClusterMA}
\BIBentryALTinterwordspacing
C.~Ding and X.~He, ``Cluster merging and splitting in hierarchical clustering algorithms,'' \emph{2002 IEEE International Conference on Data Mining, 2002. Proceedings.}, pp. 139--146, 2002. [Online]. Available: \url{https://api.semanticscholar.org/CorpusID:7563771}
\BIBentrySTDinterwordspacing

\bibitem{Deborah2015ACE}
\BIBentryALTinterwordspacing
H.~Deborah, N.~Richard, and J.~Y. Hardeberg, ``A comprehensive evaluation of spectral distance functions and metrics for hyperspectral image processing,'' \emph{IEEE Journal of Selected Topics in Applied Earth Observations and Remote Sensing}, vol.~8, pp. 3224--3234, 2015. [Online]. Available: \url{https://api.semanticscholar.org/CorpusID:1123693}
\BIBentrySTDinterwordspacing

\bibitem{Jarman2024LocalBE}
\BIBentryALTinterwordspacing
S.~Jarman, Z.~Hampel-Arias, A.~Carr, and K.~R. Moon, ``Local background estimation for improved gas plume identification in hyperspectral images,'' \emph{IGARSS 2024 - 2024 IEEE International Geoscience and Remote Sensing Symposium}, pp. 8005--8009, 2024. [Online]. Available: \url{https://api.semanticscholar.org/CorpusID:267199830}
\BIBentrySTDinterwordspacing

\bibitem{Byrd1995ALM}
\BIBentryALTinterwordspacing
R.~H. Byrd, P.~Lu, J.~Nocedal, and C.~Zhu, ``A limited memory algorithm for bound constrained optimization,'' \emph{SIAM J. Sci. Comput.}, vol.~16, pp. 1190--1208, 1995. [Online]. Available: \url{https://api.semanticscholar.org/CorpusID:6398414}
\BIBentrySTDinterwordspacing

\bibitem{Klein2022QuantifyingUI}
\BIBentryALTinterwordspacing
N.~Klein, A.~Carr, Z.~Hampel-Arias, A.~Ziemann, E.~Flynn, and K.~Mitchell, ``Quantifying uncertainty in machine learning for hyperspectral target detection and identification,'' in \emph{Defense + Commercial Sensing}, 2022. [Online]. Available: \url{https://api.semanticscholar.org/CorpusID:249260871}
\BIBentrySTDinterwordspacing

\bibitem{HampelArias20242DSR}
\BIBentryALTinterwordspacing
Z.~Hampel-Arias, A.~Carr, N.~Klein, and E.~Flynn, ``2d spectral representations and autoencoders for hyperspectral imagery classification and explanability,'' \emph{2024 IEEE Southwest Symposium on Image Analysis and Interpretation (SSIAI)}, pp. 45--48, 2024. [Online]. Available: \url{https://api.semanticscholar.org/CorpusID:269466446}
\BIBentrySTDinterwordspacing

\bibitem{HasanPour2016LetsKI}
\BIBentryALTinterwordspacing
S.~H. HasanPour, M.~Rouhani, M.~Fayyaz, and M.~Sabokrou, ``Lets keep it simple, using simple architectures to outperform deeper and more complex architectures,'' \emph{ArXiv}, vol. abs/1608.06037, 2016. [Online]. Available: \url{https://api.semanticscholar.org/CorpusID:3388909}
\BIBentrySTDinterwordspacing

\bibitem{Hu2017SqueezeandExcitationN}
\BIBentryALTinterwordspacing
J.~Hu, L.~Shen, S.~Albanie, G.~Sun, and E.~Wu, ``Squeeze-and-excitation networks,'' \emph{2018 IEEE/CVF Conference on Computer Vision and Pattern Recognition}, pp. 7132--7141, 2017. [Online]. Available: \url{https://api.semanticscholar.org/CorpusID:140309863}
\BIBentrySTDinterwordspacing

\bibitem{Kingma2014AdamAM}
\BIBentryALTinterwordspacing
D.~P. Kingma and J.~Ba, ``Adam: A method for stochastic optimization,'' \emph{CoRR}, vol. abs/1412.6980, 2014. [Online]. Available: \url{https://api.semanticscholar.org/CorpusID:6628106}
\BIBentrySTDinterwordspacing

\bibitem{Holmes2006ARO}
\BIBentryALTinterwordspacing
N.~S. Holmes and L.~Morawska, ``A review of dispersion modelling and its application to the dispersion of particles : An overview of different dispersion models available,'' \emph{Atmospheric Environment}, vol.~40, pp. 5902--5928, 2006. [Online]. Available: \url{https://api.semanticscholar.org/CorpusID:97673369}
\BIBentrySTDinterwordspacing

\bibitem{Bischl2021HyperparameterOF}
\BIBentryALTinterwordspacing
B.~Bischl, M.~Binder, M.~Lang, T.~Pielok, J.~Richter, S.~Coors, J.~Thomas, T.~Ullmann, M.~Becker, A.~Boulesteix, D.~Deng, and M.~T. Lindauer, ``Hyperparameter optimization: Foundations, algorithms, best practices, and open challenges,'' \emph{Wiley Interdisciplinary Reviews: Data Mining and Knowledge Discovery}, vol.~13, 2021. [Online]. Available: \url{https://api.semanticscholar.org/CorpusID:235829291}
\BIBentrySTDinterwordspacing

\bibitem{Akiba2019OptunaAN}
\BIBentryALTinterwordspacing
T.~Akiba, S.~Sano, T.~Yanase, T.~Ohta, and M.~Koyama, ``Optuna: A next-generation hyperparameter optimization framework,'' \emph{Proceedings of the 25th ACM SIGKDD International Conference on Knowledge Discovery \& Data Mining}, 2019. [Online]. Available: \url{https://api.semanticscholar.org/CorpusID:196194314}
\BIBentrySTDinterwordspacing

\end{thebibliography}
%



%
\newpage
\begin{IEEEbiography}[{\includegraphics[width=1in,height=1.25in,clip,keepaspectratio]{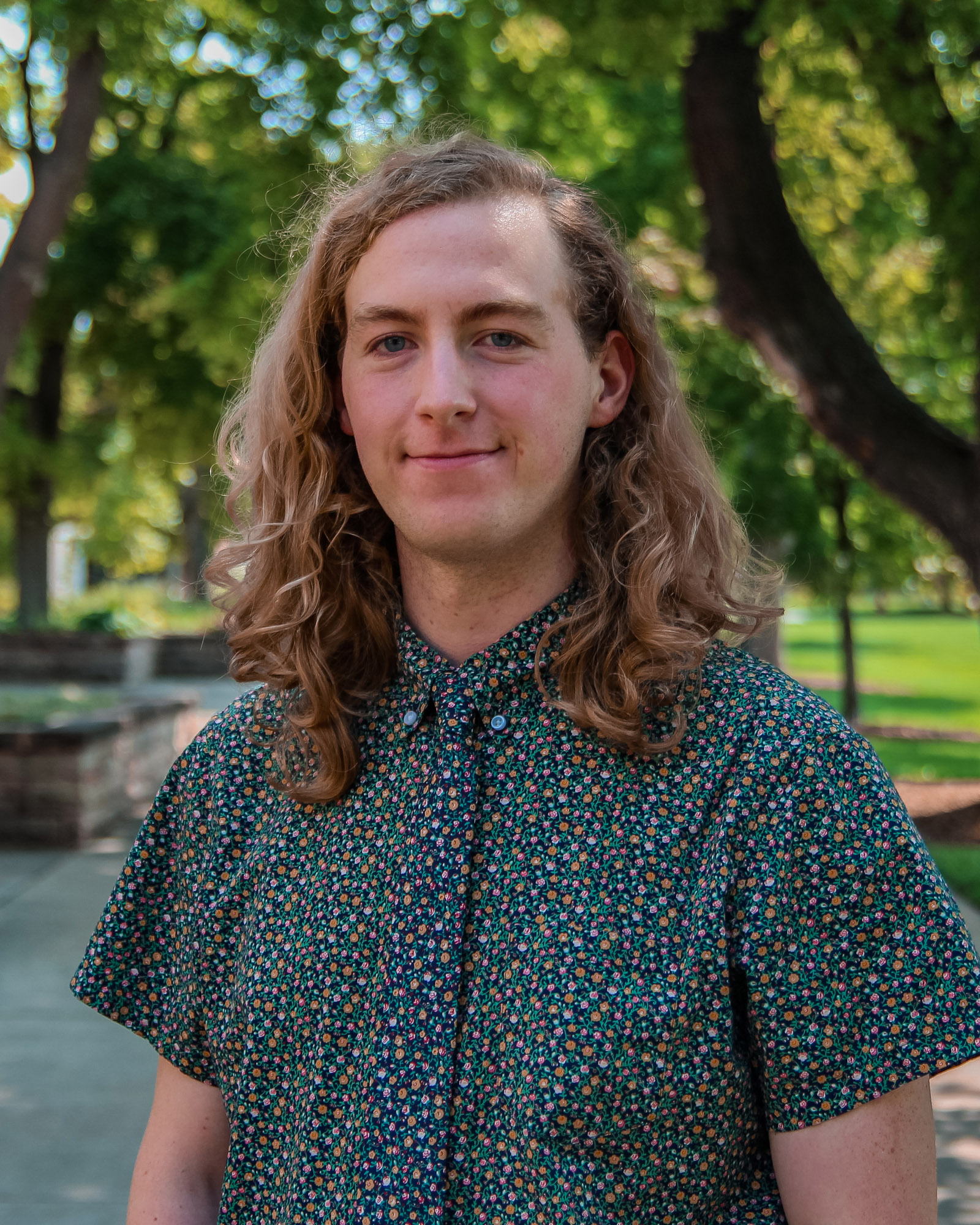}}]{Scout Jarman}
got a B.S. in Mathematics and Statistics Composite with a Computer Science minor from Utah State University in 2017.
He is currently pursuing a Ph.D. in Mathematical Sciences (emphasis in Statistics) from Utah State University under the advisorship of Dr. Kevin R. Moon, and is conducting research in collaboration with the Intelligence and Space Research group at Los Alamos National Laboratory.
His research interests are in applications of statistics and machine/deep learning to LWIR hyperspectral image analysis.
\end{IEEEbiography}

\begin{IEEEbiography}[{\includegraphics[width=1in,height=1.25in,clip,keepaspectratio]{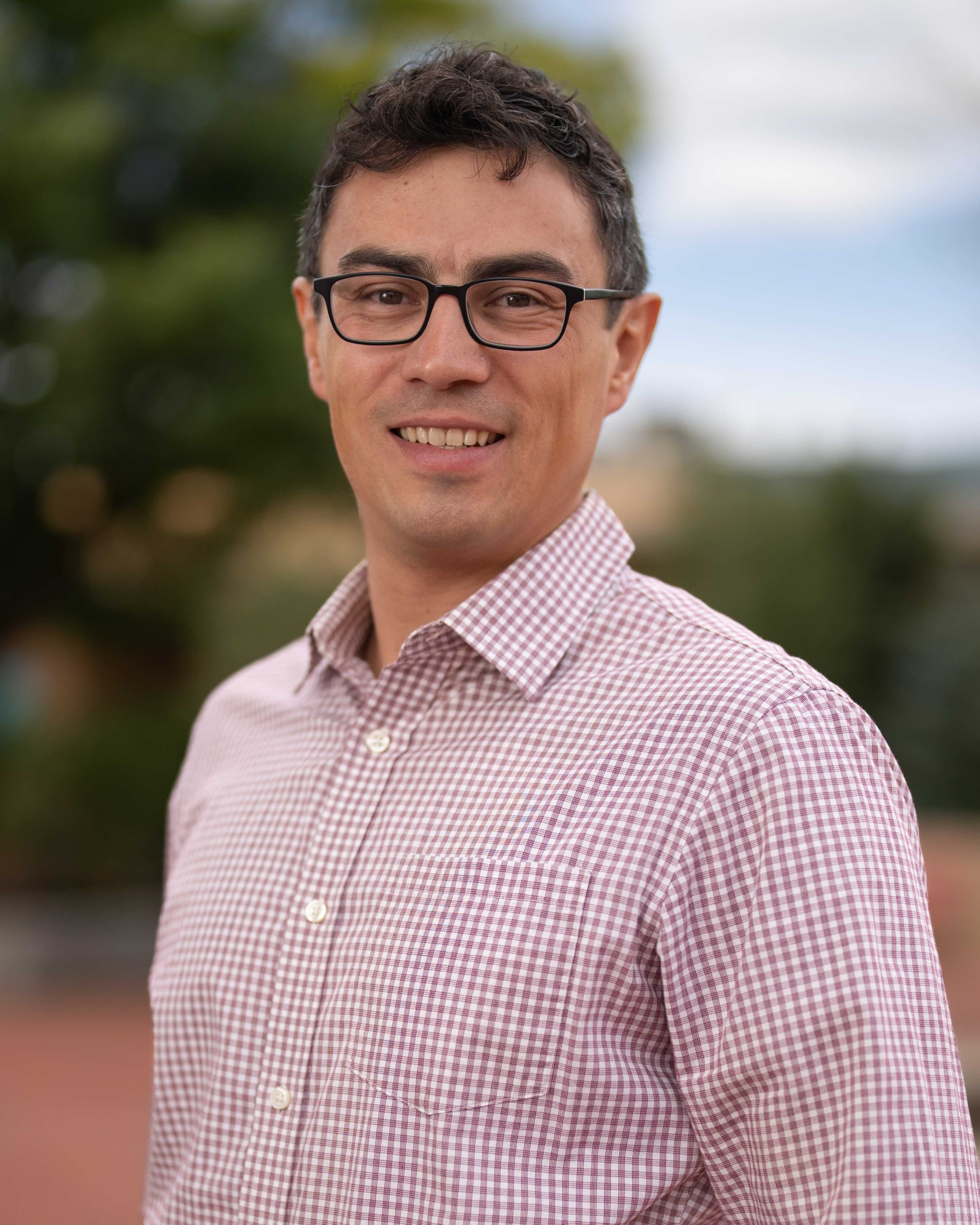}}]{Zigfried Hampel-Arias}
is a scientist in the Data Science and Remote Sensing group of the Intelligence and Space Research Division at LANL.
Dr. Hampel-Arias received his B.S. in Chemical Physics from Rice University, his Ph.D. in Particle Astrophysics from the University of Wisconsin - Madison and was a Postdoctoral Research Associate in the IIHE group at Universit\`e Libre de Bruxelles.
He has since directed his research interests in support of national security efforts using machine learning and high-performance computing.
\end{IEEEbiography}


\begin{IEEEbiography}[{\includegraphics[width=1in,height=1.25in,clip,keepaspectratio]{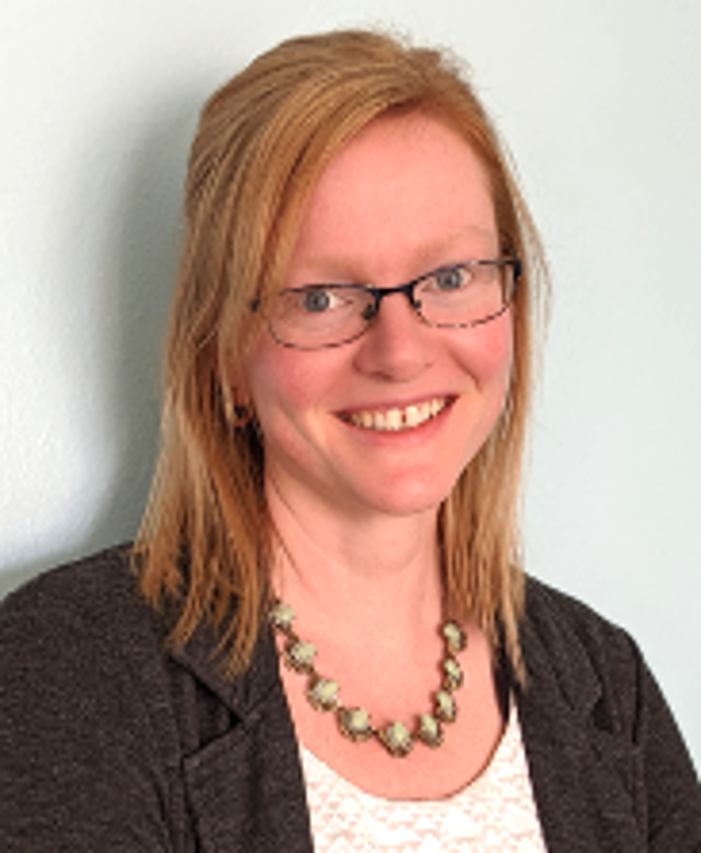}}]{Adra Carr}
holds a Ph.D in Experimental Condensed Matter Physics from University of Colorado-Boulder, and a B.S. in Physics from the University of Arizona. Her background is in ultrafast laser spectroscopy and industrial material science research, having worked on next-generation logic and memory technologies for IBM. In her current position at Los Alamos National Laboratory, her research is focused on computational imaging and deep learning applied to hyperspectral imaging.  
\end{IEEEbiography}

\begin{IEEEbiography}[{\includegraphics[width=1in,height=1.25in,clip,keepaspectratio]{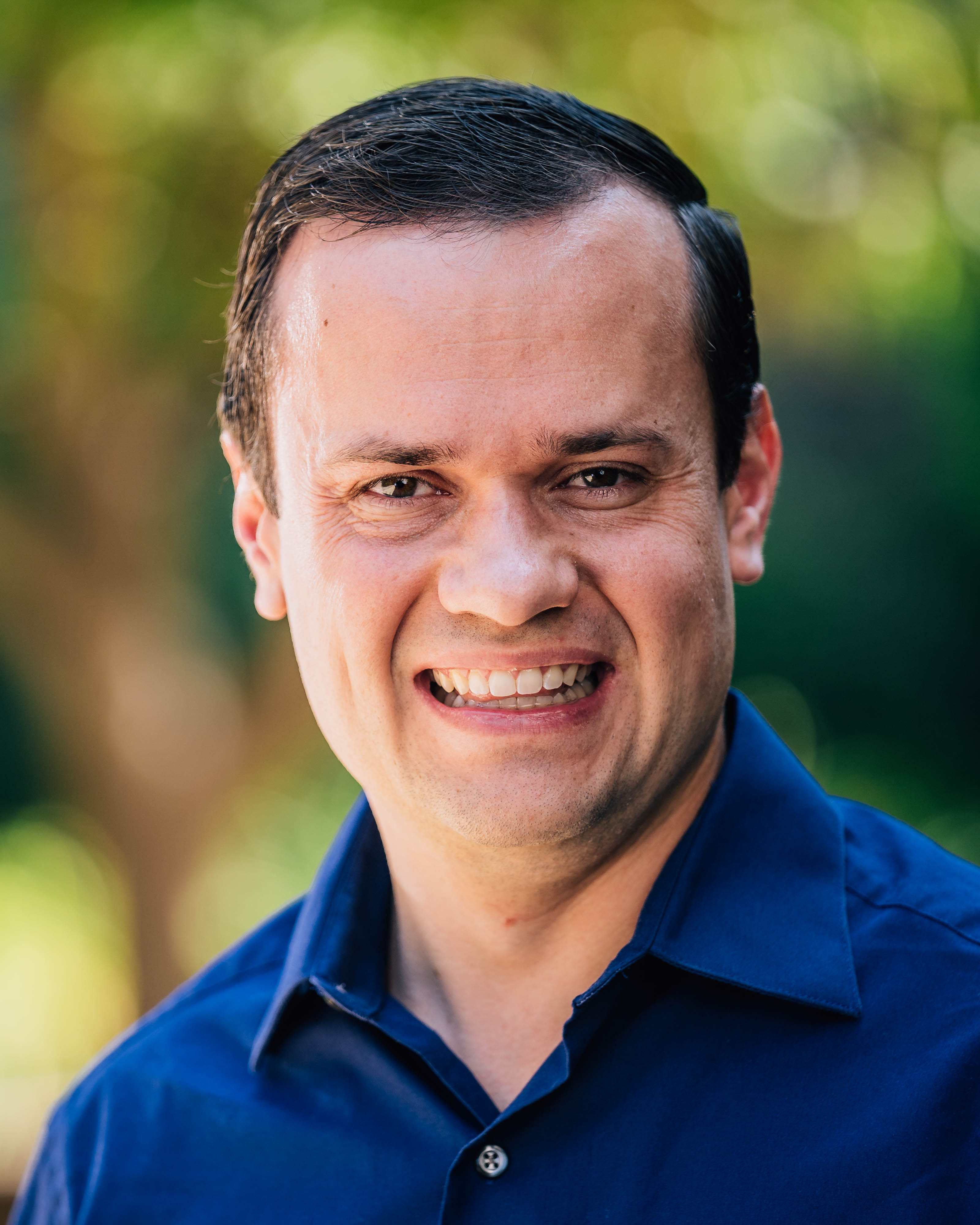}}]{Kevin R. Moon}
 is an associate professor in the Department of Mathematics and Statistics and the Director of the Data Science and Artificial Intelligence Center at Utah State University (USU). He holds a B.S. and M.S. degree in Electrical Engineering from Brigham Young University and an M.S. degree in Mathematics and a Ph.D. in Electrical Engineering from the University of Michigan. Prior to joining USU in 2018, he was a postdoctoral scholar (2016-2018) in the Genetics Department and the Applied Mathematics Program at Yale University. His research interests are in the development of theory and applications in machine learning, information theory, deep learning, and manifold learning. More specifically, he is interested in dimensionality reduction, data visualization, ensemble methods, nonparametric estimation, information theoretic functionals, neural networks, and applications in biomedical data, finance, engineering, and ecology.
\end{IEEEbiography}



\end{document}